\documentclass[12pt,preprint]{aastex}
\newcommand{\bvec}[1]{\textbf{\textit{#1}}}

\shorttitle{Weak Lensing Study of CL 0152-1357}
\shortauthors{Jee et al.}
\begin{document}
\title{Weak Lensing Analysis of the z$\simeq$0.8 cluster CL 0152-1357 with the Advanced 
Camera for Surveys}
\author{M.J. JEE\altaffilmark{1}, R.L. WHITE\altaffilmark{2},
        N. BEN{\'{I}}TEZ\altaffilmark{1},
        H.C. FORD\altaffilmark{1}, 
        J.P. BLAKESLEE\altaffilmark{1}, P. ROSATI\altaffilmark{3},
        R. DEMARCO\altaffilmark{1}, G.D. ILLINGWORTH\altaffilmark{4}}
\altaffiltext{1}{Department of Physics and Astronomy, Johns Hopkins
University, 3400 North Charles Street, Baltimore, MD 21218.}
\altaffiltext{2}{STScI, 3700 San Martin Drive, Baltimore, MD 21218.}
\altaffiltext{3}{European Southern Observatory, Karl-Schwarzschild-Strasse 2, D-85748 
Garching, Germany.}
\altaffiltext{4}{UCO/Lick Observatory, University of California, Santa
Cruz, CA 95064.}
\begin{abstract}
We present a weak lensing analysis of the X-ray luminous cluster CL 0152-1357 at $z\simeq0.84$
using HST/ACS observations. The unparalleled resolution and sensitivity of ACS enable
us to measure weakly distorted, faint background galaxies to the extent that the number
density reaches $\sim175$ $\mbox{arcmin}^{-2}$. The PSF
of ACS has a complicated shape that also varies across the field. We construct a PSF
model for ACS from an extensive investigation of 47 Tuc stars in a modestly crowded region.
We show that this model PSF excellently describes the PSF variation pattern
in the cluster observation when a slight adjustment of ellipticity is applied. The high
number density of source galaxies and the accurate removal of the PSF effect through 
moment-based deconvolution allow us to restore the dark matter distribution of the cluster in
great detail. 

The direct comparison of the mass map with the X-ray morphology from $Chandra$ observations
shows that the two peaks of intracluster medium traced by X-ray emission
are lagging behind the corresponding dark matter clumps, indicative of an on-going merger.
The overall mass profile of the cluster can be well described by an NFW profile with
a scale radius of $r_s =309\pm45$ kpc and a concentration parameter of $c=3.7\pm0.5$. 
The mass estimates from the lensing analysis are consistent with those from X-ray
and Sunyaev-Zeldovich analyses. 
The predicted velocity dispersion is also in good agreement with 
the spectroscopic measurement from VLT observations. In the adopted
cosmology where $\Omega_M = 0.27$, $\Omega_{\Lambda} = 0.73$, and $h=0.71$, the total
projected mass and the mass-to-light ratio within 1 Mpc are estimated to be (4.92 $\pm$ 0.44)$\times10^{14}M_{\sun}$ and 
95 $\pm$ 8 $M_{\sun} / L_{B\sun}$, respectively.
\end{abstract}


\section{INTRODUCTION}

Gravitational lensing has been a unique tool to probe the intervening 
matter distribution between the observer and source objects without
any assumption about the dynamical phase of deflectors. The most impressive and beautiful 
images of giant arcs can appear when the light from background sources passes near
the caustic of a foreground lens. These ``strong lensing'' features
indicate the presence of a critical (or higher) surface mass density in the inner region 
and are particularly useful
for constraining the core structure of the lensing cluster.
In the ``weak lensing'' regime where the distortion becomes weaker and
less obvious (whether the cluster is less massive or projected distances of source galaxies are farther from the cluster center), 
one can still detect coherent alignments of background galaxies and restore
the mass distribution up to far greater radii from these subtle measurements.
Since the first successful detection of systematic alignments of background galaxies 
by \citet{twv}, the technique has been 
applied to a wide selection of galaxy clusters and is now firmly established
as one of the most straightforward paths to probe the mass distribution of the cluster in question.

In general, the detectability of weak gravitational shears depends not only upon
the intrinsic signal strength determined by the projected mass density of a lens and the geometry 
between lens and source, but also upon the observational restrictions
set by finite sensitivity and resolution of an instrument. In this regard, a weak lensing
analysis of high-redshift clusters is disadvantaged because the signal
decreases as the redshift of the lensing cluster approaches that of background galaxies, and
also it becomes harder to recover shapes of faint, poorly resolved galaxies, which however contain
most of the useful signal. Nevertheless, the demands for comprehensive studies on 
many individual high-redshift clusters are increasing because of their potentially significant implications
for cluster formation and cosmology. For example, even the mere abundance of massive
clusters at such high redshifts can strongly constrain the cosmological density parameter $\Omega$,
decoupling the $\sigma_8 \Omega ^{0.5} \simeq0.5$ degeneracy \citep[e.g.,][]{cmye,bf}.
Furthermore, most high-redshift clusters possess filamentary structures indicating their early stage of formation, and
it is interesting to investigate in detail how dark matter is distributed with respect to cluster
galaxies or the intracluster medium (ICM) traced by X-ray emission. 

The remarkable substructure of MS 1054-03 obtained through the weak lensing analysis of WFPC2 observations
\citep[hereafter HFK00]{hfk00} demonstrated the undeniable merits of space-based
weak lensing observations and already hinted at the bright prospects of the newly installed Advanced Camera 
for Surveys (ACS) in this application. The advantages of HST observations over ground-based imaging include 
a higher number density of background galaxies whose shapes can be reliably determined,
and smaller corrections for point spread function (PSF) effects.
The pixel size of the Wide Field Channel of ACS is $\sim$0.05 $\arcsec$, offering twice the sampling resolution 
of the Wide Field (WF) chips of WFPC2. In addition, ACS has a factor of 5 improvement in throughput while
providing twice the field of view of WFPC2. Therefore, the finer resolution, higher
sensitivity, and wider field of view of ACS can provide a much higher fraction of well-resolved galaxies
with less investment of HST observing time. 

In this paper, we present a weak lensing study of the X-ray selected high-redshift (z$\sim$0.84)
cluster CL 0152-1357 using the Wide Field Channel of ACS. Together with
MS 1054-03, CL 0152-1357 is one of the most X-ray luminous clusters at $z \ge 0.8$ known to date
whose X-ray properties have been well-studied by the ROSAT \citep{ebeling00}, $BeppoSAX$ \citep{della00},
and $Chandra$ Observatory \citep{maughan03}. Nevertheless, unlike MS 1054-03, there have been no 
HST-based high-resolution weak lensing studies so far. Such studies, combined with X-ray observations, can 
substantially enhance our understanding of the dynamical evolution of the ICM and
its interaction with cluster galaxies as well as of the cluster substructure as a whole. Particularly, CL 0152-1357
is known to have a complicated X-ray and optical substructure, and the high-resolution mass reconstruction
from the ACS weak lensing is expected to provide the mass distribution of the cluster in unprecedented detail.

We organize our works as follows. In \textsection\ref{obs}, we describe the observations and basic data reduction. 
Ellipticity measurement of source galaxies is presented in \textsection\ref{elle}. In \textsection\ref{psf}
we discuss the PSF modeling and correction. \textsection\ref{lum} describes the luminosity of the cluster and the
distribution thereof. The mass reconstruction of the cluster is handled in \textsection\ref{massrecon_section} and the total
mass estimates from various approaches are described in \textsection\ref{massest}. Finally, the substructure 
from this study is compared with that of other studies in \textsection\ref{substructure} before the conclusion \textsection\ref{conclusion}.
Throughout the work we assume $\Lambda$ CDM cosmology favored by the Wilkinson Microwave Anisotropy Probe (WMAP) with $\Omega_M = 0.27$, $\Omega_{\Lambda}=0.73$, and
$H_0 = 71 \mbox{km s}^{-1} \mbox{Mpc}^{-1}$. 

\section{OBSERVATIONS\label{obs}} 
 
\begin{figure}
\plotone{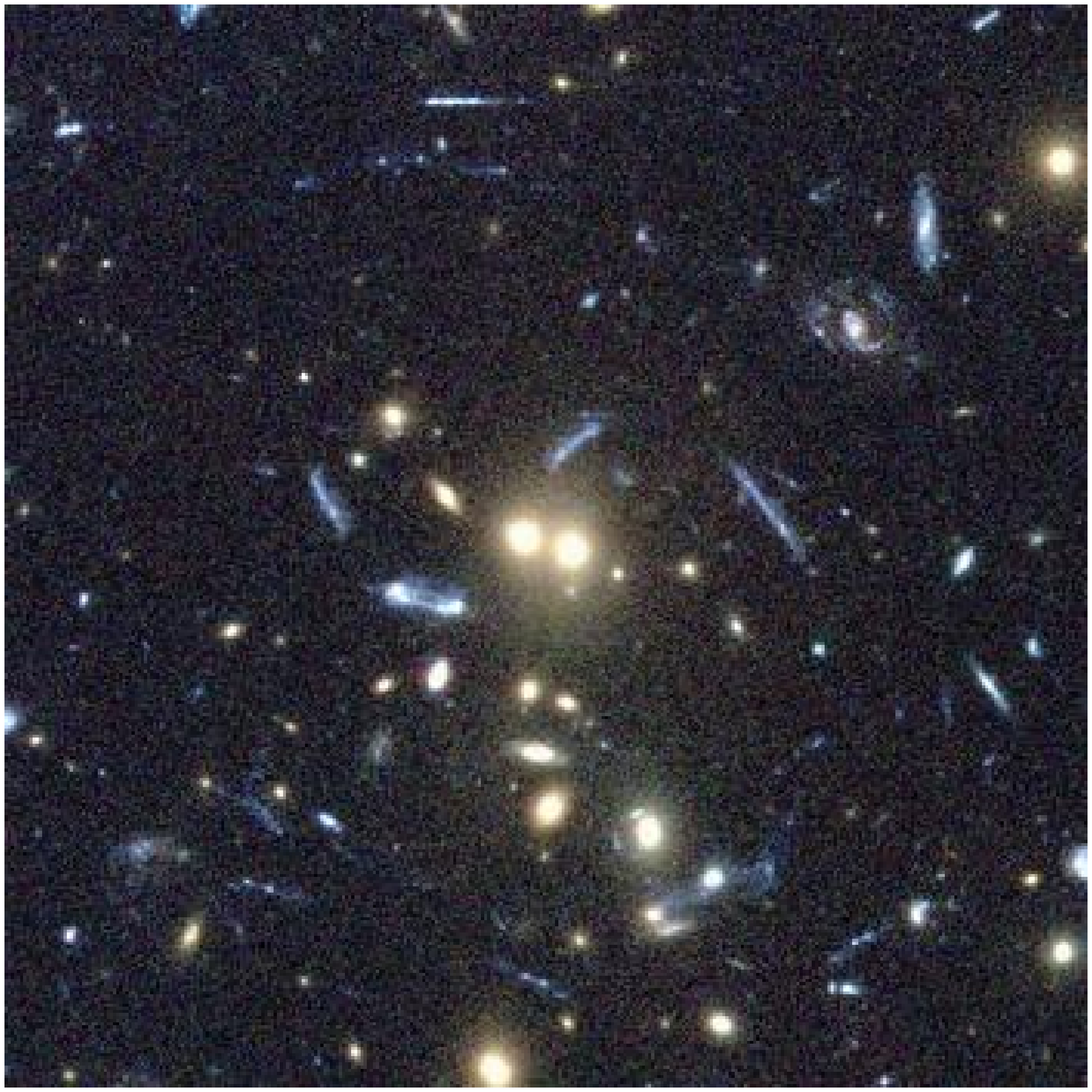}
\caption{Color composite of CL 0152-1357 center. The image shows the central $\sim40\arcsec \times 40\arcsec$ section
of the entire $\sim350\arcsec \times 350\arcsec$ field. The image is created 
using the FITSCUT \citep{mccann04}.  \label{fig1}}
\end{figure}

\begin{figure}
\plotone{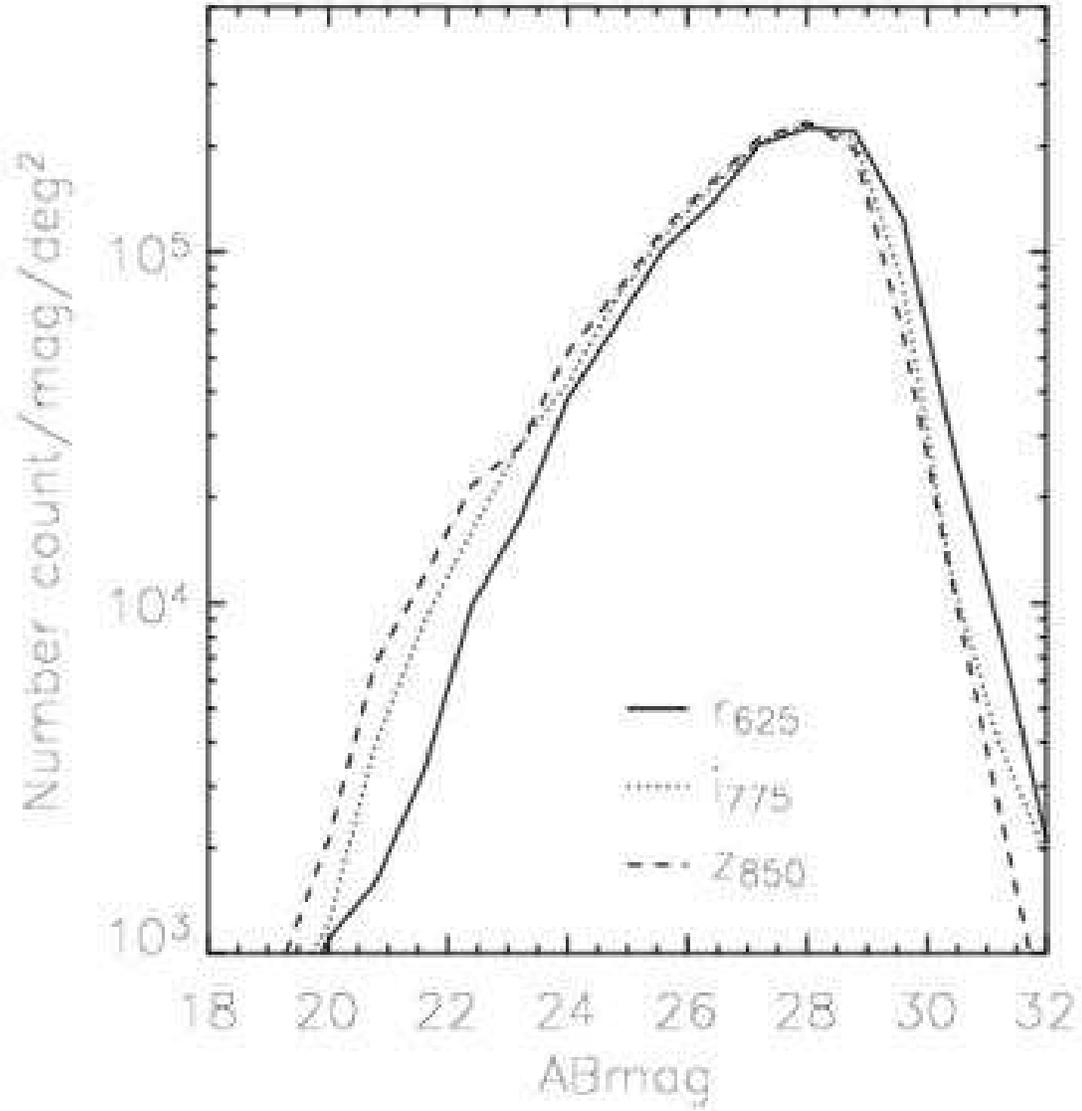}
\caption{Number counts of galaxies in the $r_{625}$ (solid), $i_{775}$ (dotted), and $z_{850}$ (dashed) passbands. The
detection is complete down to $\sim27.5$ mag in all filters.  \label{fig2}}
\end{figure}

CL 0152-1357 was observed in a 2$\times$2 mosaic pattern allowing
$\sim50\arcsec$ overlap between pointings with
the Wide Field Channel of ACS during 2002 November and 2002 December (GTO proposal 9290, P.I. Ford). 
The cluster was imaged in F625W, F775W, and F850LP  (hereafter $r_{625}, i_{775},$ and $z_{850}$, respectively) with integrated
exposure per pointing $\sim4800$ s. 
The low level CCD processing (e.g. overscan, bias, dark subtraction, and 
flat-fielding) was done with the standard STScI CALACS pipeline, and the final high-level
science images were created through the ``apsis'' ACS GTO pipeline \citep{blakeslee03}.
The resulting high accuracy in both the image registration and the geometric distortion correction
is indispensable in weak lensing measurements. The apsis pipeline calculates offsets and rotations 
between images using the ``match'' program \citep{richmond02} after applying a geometric distortion model \citep{meurer03}
to the astronomical objects which have a high signal-to-noise ratio (SNR).
The typical shift uncertainty is $\sim$ 0.015 pixels (J. Blakeslee 2004, private communication).
The drizzle-blot-drizzle cycle
of apsis automatically rejects cosmic rays and generates mosaic science images. We used
the Lanczos3 drizzling kernel, which gives a sharper PSF and
less noise correlation between neighboring pixels. The noise correlation decreases the root mean square
(RMS) noise fluctuations and therefore causes photometric errors to be slightly underestimated.
Apsis calculates the correct RMS noise in the absence
of correlation. The RMS map produced in this way is used for source detection and
photometric error estimation. We present the color composite image of the cluster center
in Figure~\ref{fig1}, which shows many strong-lensing features such as arc(let)s around
the two central brightest cluster galaxies (BCGs).
The objects were detected using SExtractor \citep{ba96} on the detection image created
by combining all the present filter images after applying inverse variance weighting. In general, many
parameters in SExtractor affect the detection procedure, and we adopted the values
obtained from the experiments of \citet{benitez04}. They tried to suppress 
spurious detections while extracting all obvious galaxies by tuning up these parameters. We
found that their rather conservative choice of {\tt DETECT\_MINAREA}=5 and {\tt DETECT\_THRESH}=1.5 selects
faint galaxies up to the detection limit (S/N $\sim$ 3) without significantly introducing
false objects into the object catalog.
We ran SExtractor in dual image mode for each filter to obtain the galaxy photometry and
colors. The final catalogs were visually compared with the image in order to remove false detections (e.g. 
diffraction spikes and spurious spots around bright stars, saturated CCD bleeding, noise fluctuation at image edges, 
clipped and merged galaxies, HII regions inside nearby galaxies, etc.). 
This resulted in a total of 10992 objects. The number count plot in Figure~\ref{fig2}
shows the completeness of our data down to $\sim$27.5 magnitude in all three passbands.

\section{ELLIPTICITY MEASUREMENTS\label{elle}}

In the regime where the background galaxy is much smaller than the scale length
of the gravitational potential variation, we can obtain the linearized lens mapping equation as follows:
\begin{equation}
\textbf{A}(\bvec{x}) =  \delta_{ij} - \frac {\partial^2 \Psi (\bvec{x})} {\partial x_i \partial x_j}
= \left ( \begin{array} {c c} 1 - \kappa - \gamma _1 & -\gamma _2 \\
                      -\gamma _2 & 1- \kappa + \gamma _1  
	  \end{array}   \right ),  \label{equ_lens}
\end{equation} 
\noindent
where $\textbf{A}(\bvec{x})$ is the transformation matrix $\bvec{x}^\prime = \textbf{A} \bvec{x}$ which
relates a position $\bvec{x}$ in source plane to a position $\bvec{x}^\prime$ in image plane,
and $\Psi$ is the 2-dimensional
lensing potential. In the matrix of equation~\ref{equ_lens}, the convergence $\kappa$ determines the overall magnification, and
$\gamma_1$ and $\gamma_2$ describe the shear along x-axis and at $45^{\degr}$ from the x-axis,
respectively. Therefore, in general, the galaxy images are distorted in shape ($\gamma_1$, $\gamma_2$) and
size ($\kappa$) (in a strict sense, $\gamma_{1(2)}$ also alters the object size anisotropically).
Though \citet{btp95} claimed that the magnification bias (and thus the number density bias) can be used to estimate 
the local surface mass density directly, most weak lensing works have been based on the 
ellipticity biases. This is because the magnification effect is more sensitive to 
shot noise, and the SNR of the shear measurement is considerably better than
that of the magnification effect in a typical weak lensing ($\kappa \ll 1$) regime \citep{ske00}.

The ellipticity of an object can be defined in terms of weighted quadrupole moments as follows:
\begin{equation}
e = \left (  \frac {I_{11} - I_{22}}{I_{11} + I_{22}} , \frac {2 I_{12}}{I_{11}+I_{22}} \right )
\end{equation}

\begin{equation}
I_{ij}= \int w(\bvec{x}) f( \bvec{x}) x_i x_j d^2 \bvec{x},
\end{equation}
\noindent
where $f(\bvec{x})$ is the pixel intensity, and  $w(\bvec{x})$ is 
the weight function required to suppress the noise in the outer region of the object.
\citet[hereafter KSB]{ksb95} used the circular Gaussian weight whose size matches
that of the object, thus maximizing the significance of the measurement.
However, the circular weight makes the object rounder, and the effect becomes severe for
highly non-circular galaxies. Besides, the ellipticities calculated in this way do not
follow the simple ellipticity transformation rule \citep{kochanek90,miralda91} in response to
the applied shear. These features necessitate the introduction of an additional parameter
which in general depends on the higher moments. In KSB work, this quantity 
is referred to as ``shear polarizability'' $P_\gamma$.

Recently, \citet[hereafter BJ02]{bj02} introduced adaptive moments using an elliptical
weight function whose shape and size match those of an object. While the concept
of finding the optimal elliptical Gaussian weight function is mathematically simple, 
the actual implementation can take various forms. For example, one can determine
the weight function by minimizing the deviation from the image in the least-square sense. 
Alternatively, one can start with a circular weight function and iteratively
modify the ellipticity, the size, and the centroid of the weight function until these parameters converge.
BJ02 effected the determination of optimal elliptical weight function 
by iteratively shearing the objects to match the $circular$ Gaussian weight. Considering the finite pixelization 
of object images, this may not sound more attractive than the previous two schemes. However, if the galaxy images
can be decomposed via mathematically well-behaved basis functions, the adaptive elliptical
moments are computed inexpensively. BJ02 proposed the polar eigenfunctions of 2-dimensional quantum
harmonic oscillator (QHO) as basis functions. This decomposition was also independently
suggested by \citet[hereafter R03]{refregier03} though his shear estimator is different from that of BJ02.
Many mathematically convenient formalisms developed for these eigenfunctions or $shapelets$ include operators
which can effect coordinate transformations such as shear, translation, dilation, rotation, etc.
Even more important advantages one obtains from the galaxy expansion
using shapelets are that the PSF can be compactly described by the coefficients
of the basis functions and the (de)convolution is easily achieved by simple matrix manipulations.

Shapelets in polar coordinate are given by

\begin{equation}
I(r,\theta) = \sum_{p,q \ge 0} b_{pq} \Psi_{pq}^\sigma (r,\theta)
\end{equation}

\begin{equation}
\Psi_{pq}^\sigma (r,\theta) = \frac {(-1)^q }{ \sqrt { \pi} \sigma^2} \sqrt { \frac{q!}{p!}}
   \left ( \frac{r}{\sigma} \right ) ^{(p-q)} e ^{i (p-q) \theta} e^{-r^2 / 2 \sigma ^2} L_q ^{(p-q)} 
     ( \frac{r^2}{\sigma^2}) \phantom{xxxxxx} (p \geq q),
\end{equation}
\noindent
where $L_q^{m} (x)$ are the Laguerre polynomials. The complex conjugate relation
$\Psi_{qp}^\sigma = \bar{\Psi}_{qp}^\sigma$ is used to compute the eigenfunctions when $p < q$.
Other useful mathematical properties of the above basis functions along with recursion 
relations of matrix elements of the aforementioned operators are presented in BJ02.

\begin{figure}
\plotone{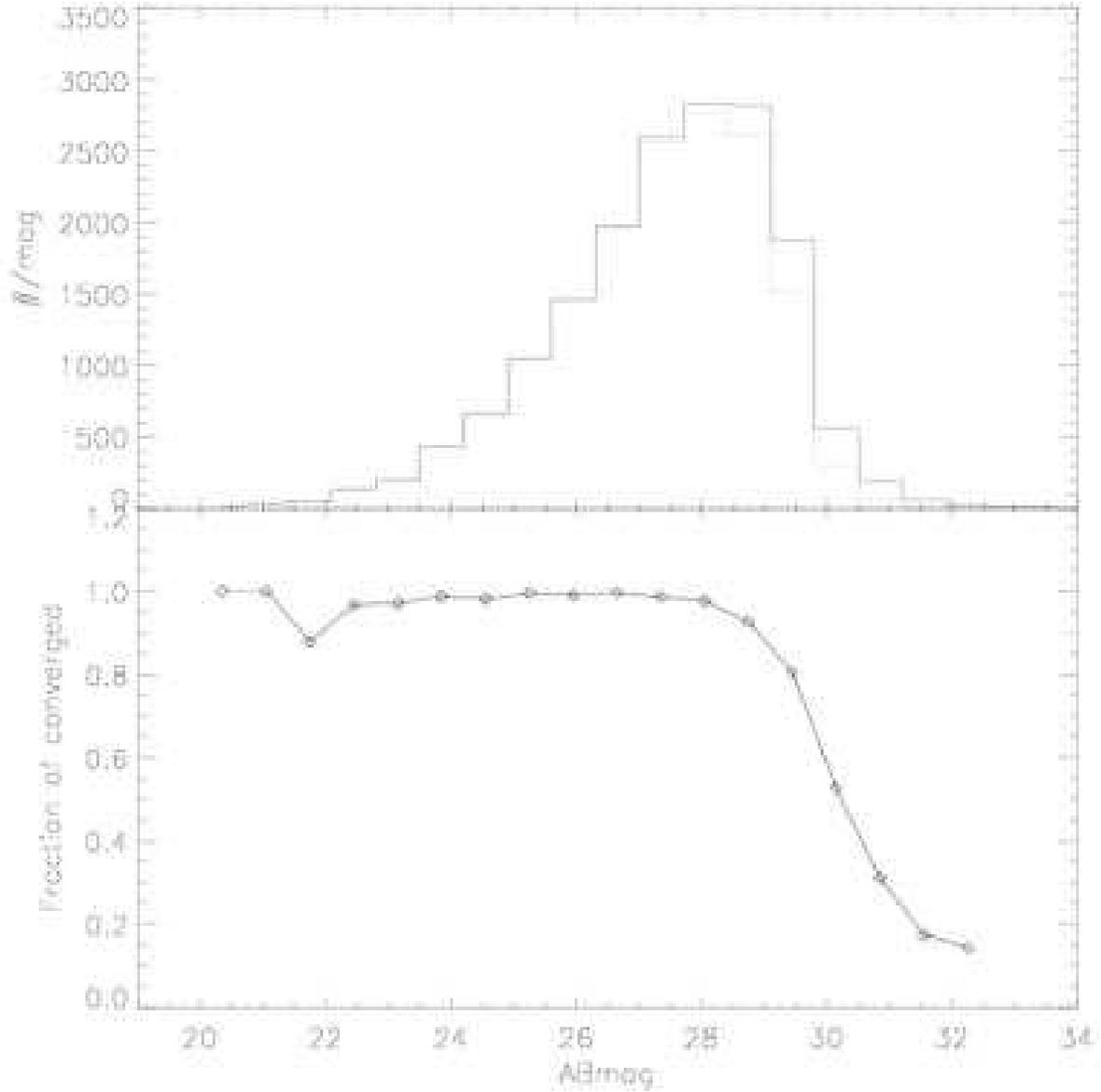}
\caption{Statistics of galaxies whose shapes are reliably measured.  Solid line (top panel)
represents the number of objects detected by SExtractor and dashed line 
indicates the number of objects for which the iteration converges. The fraction of measureable
galaxies decreases substantially after $\sim29$ mag (see bottom panel). \label{fig2.5}}
\end{figure}

Once the galaxy image is decomposed into $b_{pq}$ vectors, we can translate, dilate, and shear

\begin{equation}
\bvec{b} ^ \prime = ( \mbox{\bf{S}} _\eta \mbox{\bf{D}} _\mu \mbox{\bf{T}}  _z ) \cdot  \bvec{b}, \label{eta}
\end{equation} 
until the series of transformation satisfies the following conditions

\begin{equation}
b^\prime_{10}=0 \label{eqn_cen}
\end{equation}
\begin{equation}
b^\prime_{11}=0 \label{eqn_sig}
\end{equation}
\begin{equation}
b^\prime_{20}=0 \label{eqn_ell}.
\end{equation}
\noindent
The condition imposed by equations~\ref{eqn_cen}, \ref{eqn_sig}, and \ref{eqn_ell} relate to centroid, size, and ellipticity of the
optimal elliptical Gaussian, respectively. In this paper, the algorithm of BJ02 method is 
independently implemented in the Interactive Data Language (IDL). Figure~\ref{fig2.5} demonstrates
the statistics of the decomposition of galaxies in $r_{625}$ passband. The fraction of reliably measurable
galaxies decreases substantially if target galaxies are fainter than $r_{625}\sim29$. A similar trend
is observed in the other two passbands. The final shape catalog is produced after optimally
combining ellipticities in all three passbands.

One must note that the quantity $\eta$ in equation~\ref{eta}
is different from the conventional definitions of ellipticities though they are related in a straightforward manner.

\begin{equation}
\delta \equiv \frac {1-q^2}{1+q^2} = \tanh \eta  \label{eqn_delta}
\end{equation}

\begin{equation}
\epsilon \equiv \frac {1-q}{1+q}= \tanh \frac{\eta}{2},
\end{equation}

\noindent 
where $q$ is the axis ratio $b/a$.
In this paper, the $distortion$ $\delta$ in equation~\ref{eqn_delta} is referred to as ellipticity unless indicated otherwise.

\section{PSF CORRECTIONS \label{psf}}

As one probes to weaker and weaker lensing regimes, the
accurate removal of any instrumental artifact becomes paramount to the success of the analysis.
Finite seeing causes the cicularization of small galaxies while the anisotropy of the PSF 
can create systematic biases in the size and the direction of the polarization.
The pioneering investigation of KSB95 suggested that an approximation can be made
by treating the real PSF as a small perturbation to an isotropic PSF. The original prescription
and its variations have been employed widely during the last decade 
though the valid regime of its application was frequently questioned. \citet{kaiser00} argued
that despite the fact that the modified KSB works reasonably in some cases, the
technique may become problematic in diffraction-limited observations. 
In the current investigation, among many recent suggestions \citep[e.g.,][]{wcf96,kuijken99,kaiser00}, we 
settled upon the moment-based deconvolution technique (BJ02; R03) which
performs the deconvolution by the matrix manipulation of shapelet components of the galaxy and PSF.
The decomposition of the PSF in this way not only eases the modeling of the PSF variation across the field, but
also effectively suppresses the noise amplication if the truncation of the higher order moments is carefully
handled. 

\subsection{PSF Modeling of WFC}

\begin{figure}
\plottwo{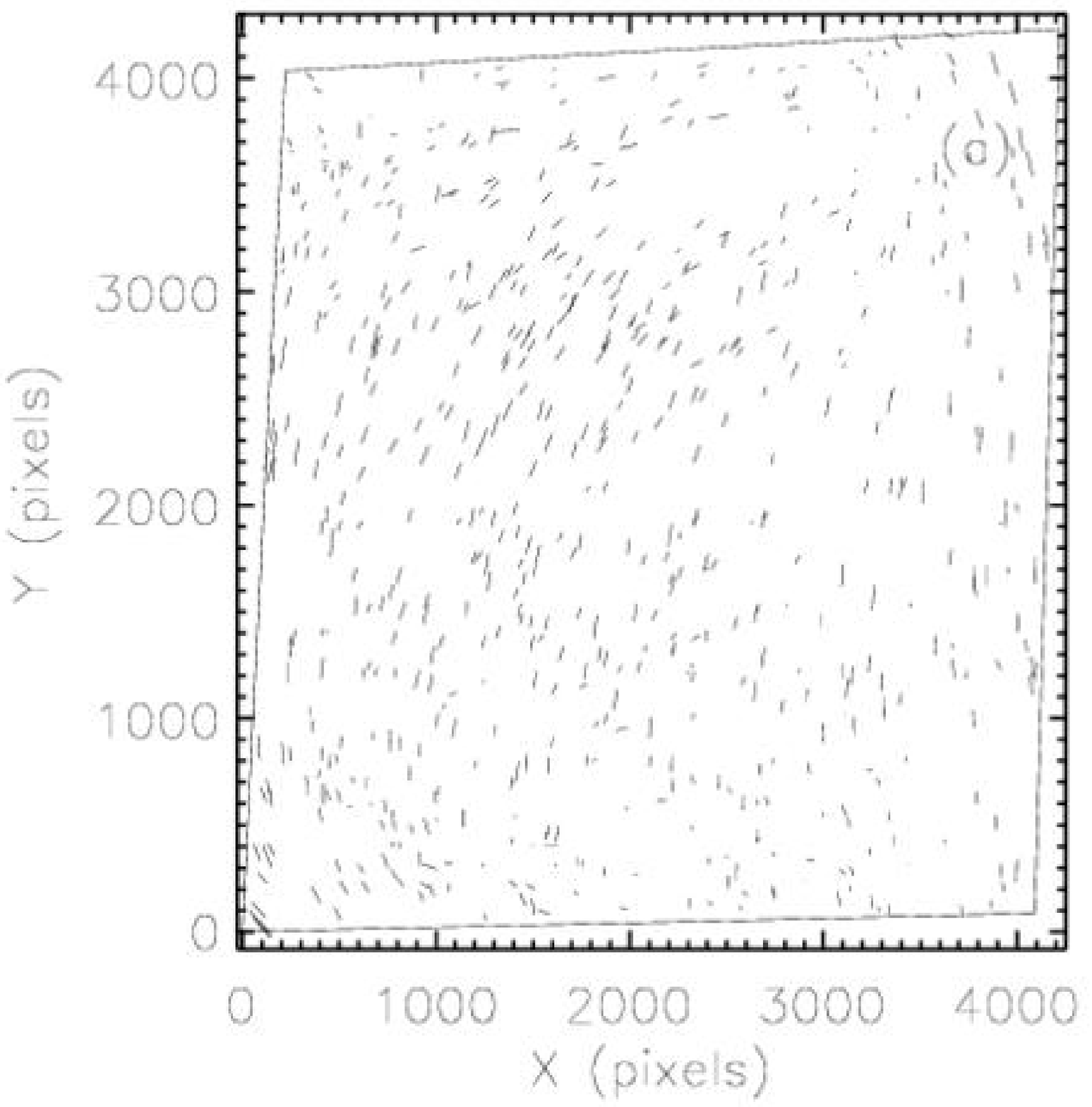}{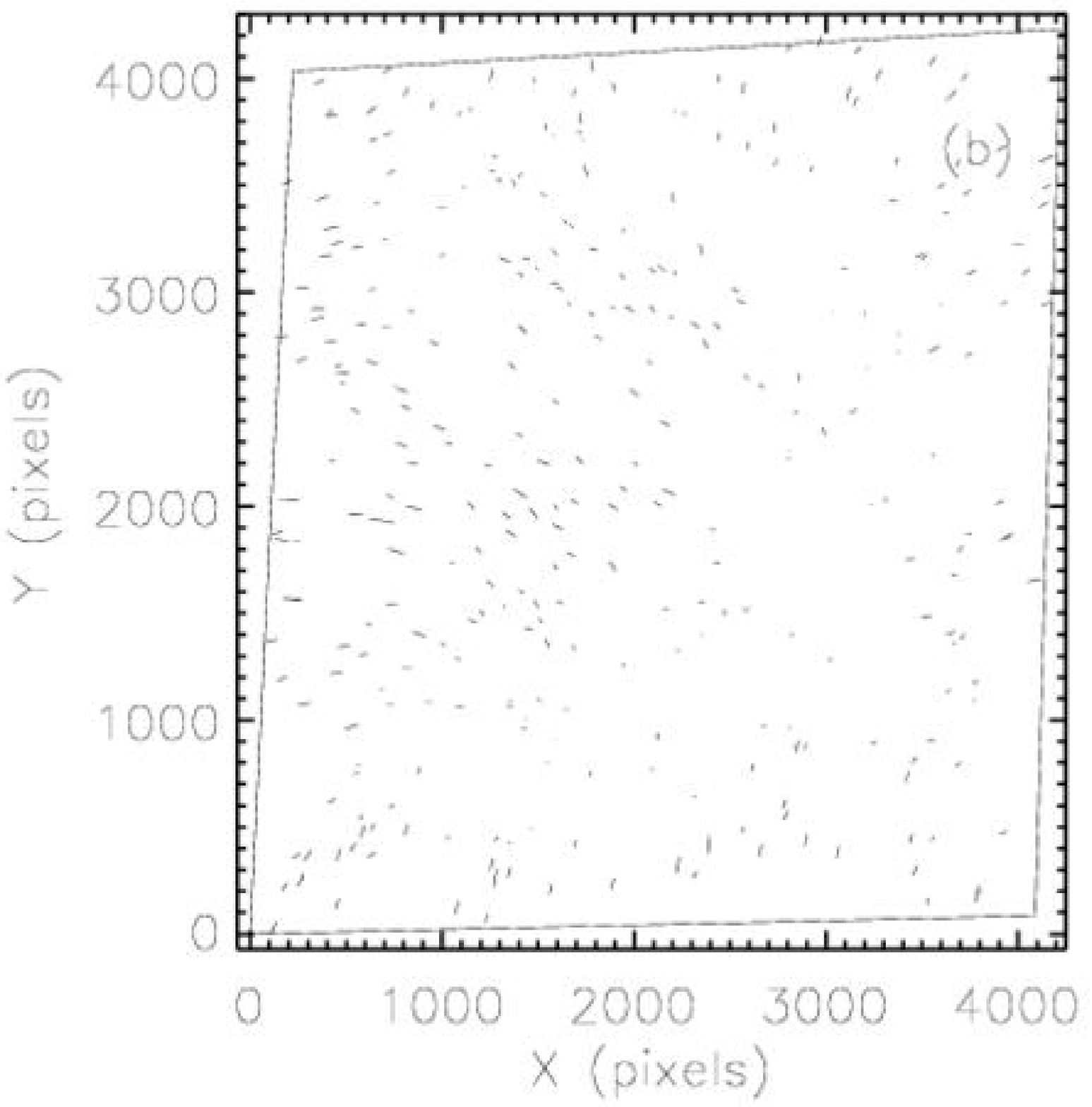}
\caption{PSF variation of WFC sampled from external field observation of 47 Tuc. 
The length and direction of the whisker are proportional to the magnitude and
orientation angle of ellipticity, respectively.
(a) PSF pattern observed on 2002 October 3. Most stars are elongated from lower-left to upper-right.; (b) The same field is taken on
2002 October 24. The PSF elongation is nearly perpendicular to the pattern in (a).  \label{fig3}}
\end{figure}

Though field-dependent variation of the WFC PSF is small compared to that of WFPC2, its change in
ellipticity within the field is significant \citep{krist03}. In order to investigate the issue
we retrieved the repeated ($\sim$ every three weeks) observations of the modestly crowded region of the globular 
cluster 47 Tuc originally used to monitor the flat-fielding stability of 
ACS (PROP 9656, PI De Marchi). Because the default drizzling kernel of the STScI pipeline is square,
we had to re-drizzle all the flat-fielded (FLT) images using Lanzcos3 kernel to match the PSF size of the CL 0152-1357 observation.
After the initial detection of stars by SExtractor, we selected ``good'' stars which are bright,
unsaturated, and isolated (having no companion stars or cosmic-rays within 15 pixels from the center).
Then, each star is decomposed into shapelet components by minimizing the following:

\begin{equation}
\chi^2 = \sum_{i} { \frac {\left ( I_i - \sum_{pq} b_{pq} \Psi ( \bvec{x} _i ) \right ) ^2 }
                     { \sigma_i^2 } },
\end{equation}

\noindent
where the optimal centroid and size of the eigenfunctions $\Psi _{pq} $ are iteratively determined.
The two whisker plots in Figure~\ref{fig3} show typical PSF patterns of the WFC measured from the 47 Tuc field.
\citet{krist03} pointed out the magnitude of ellipticity is determined by the focus offset, and
the angle of elongation switches by 90$\degr$ when the sign of the offset becomes opposite. This fact is confirmed
by comparing Figure~\ref{fig3}a and~\ref{fig3}b which are taken on 2002 October 3 and 2002 October 24, respectively, roughly
on the same patch of the 47 Tuc field. We find that 
ACS PSF patterns at other epochs follow one of these two patterns with slightly altered ellipticities.

The spatial variation of shapelet coefficients $b_{pq}$ of ACS PSFs is modeled as

\begin{equation}
b_{pq}^{\prime}  = a_{00} + a_{10} x + a_{01} y + a_{20} x^2 + a_{11} xy + a_{02} y^2 +  \cdot \cdot \cdot .
\end{equation}
\noindent

We found the third order in $x^i y^j$ (i.e. $i+j \le 3$) is sufficient to describe the pattern and higher
order polynomials do not improve the agreement between the model and the data. 
Now, the important question is how well the PSF taken from the 47 Tuc field can describe the PSF on
the CL 0152-1357 field. Since the ellipticity of the PSF changes with respect to the focus offset,
it is necessary that the ellipticity of the model PSF is made adjustable to match that of the
actual PSF in the cluster observation.
We implemented this by applying a shear operator to the shapelet
components of the model PSF. That is,

\begin{equation}
b_{pq}^{\prime} = \mbox{\bf{S}}_{\delta \eta} b_{pq},
\end{equation}
\noindent
where the evaluation of matrix elements of the shear operator $\mbox{\bf{S}} _ {\delta\eta}$ can be found in Appendix A.3 of BJ02.
Though $\delta\eta$ can be also allowed to vary depending on the position in principle, 
we found that a simple fixed parameter per exposure nicely reduce the systematic residuals.
Because we measure galaxy shapes on the $2\times2$ mosaic image, another slight complexity arises due to the overlap between pointings.
Nevertheless, a reasonable assumption can be made that the PSF in the overlapping region is very closely
approximated as an exposure-time-weighted average of all the contributing PSFs. 

\begin{figure}
\plottwo{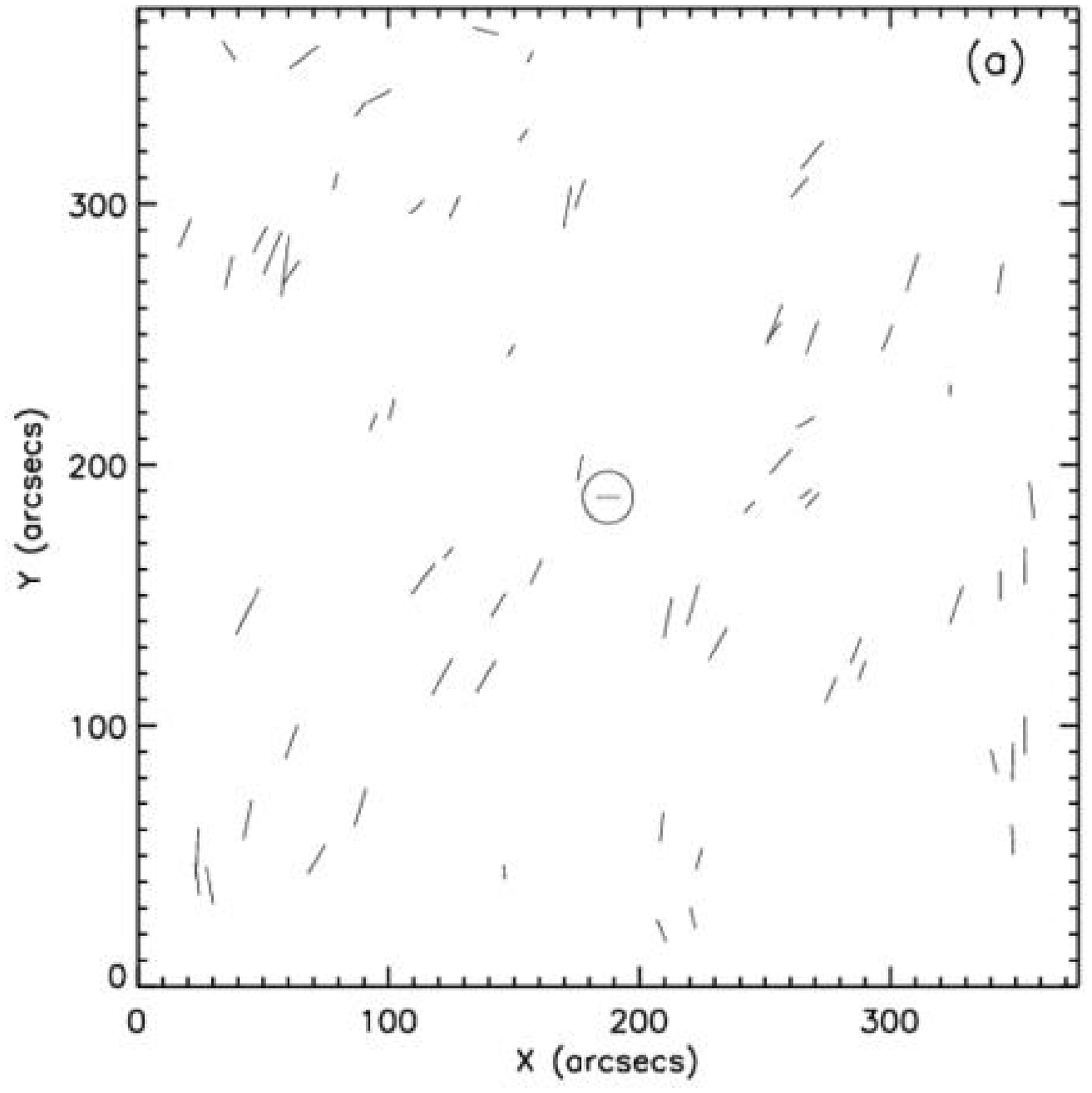}{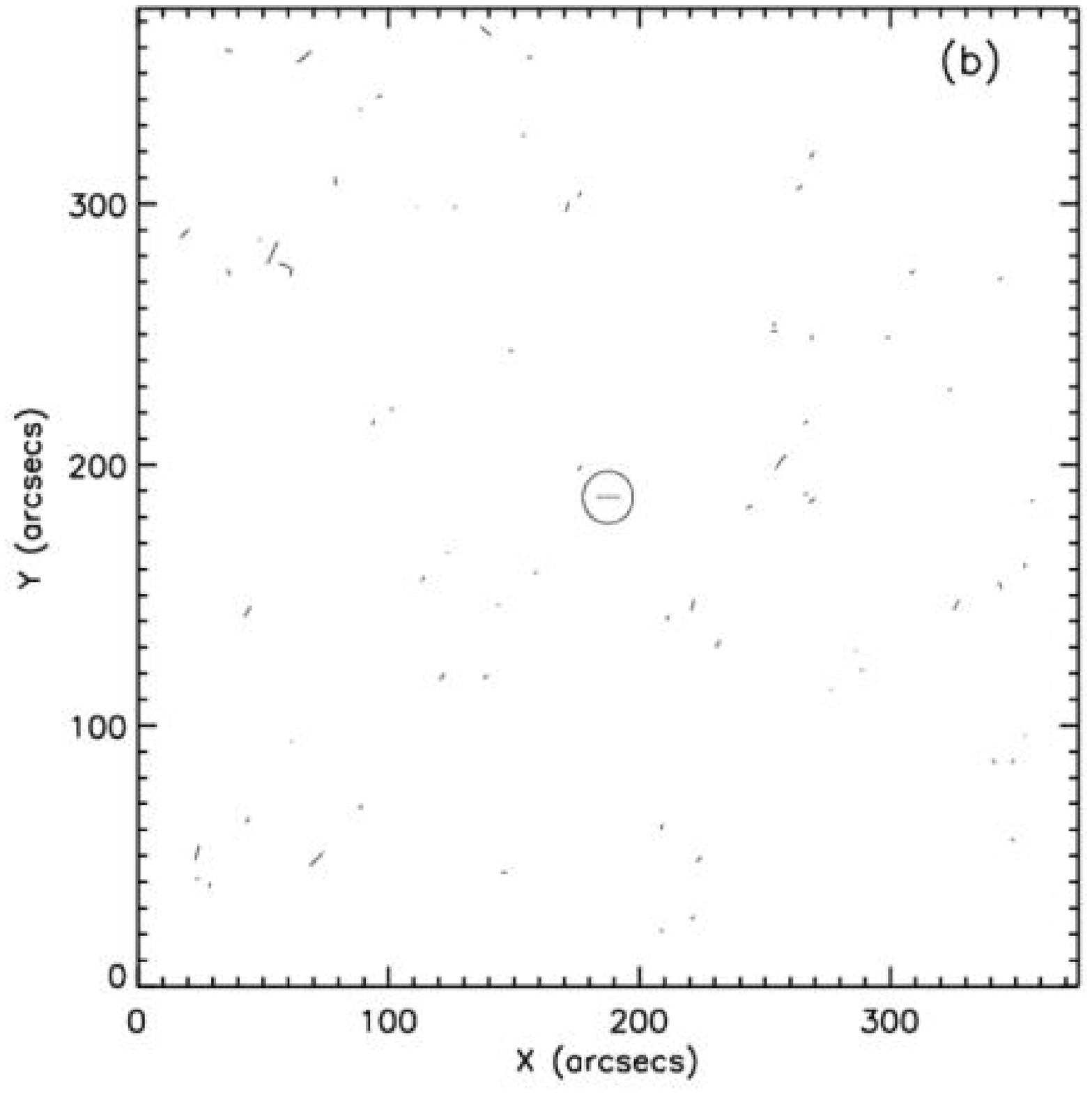}
\caption{PSF pattern in $r_{625}$ passband of CL 0152-1357 field. (a) Uncorrected ellipticities measured from stars, $ \left <
 \delta \right > = 0.114 \pm 0.040$; (b) Ellipticities after removal of PSF anisotropy 
 $\left < \delta \right > = 0.023 \pm 0.017$. Correction is made by reconvolving the
 image with position-dependent rounding kernels. The circled whiskers at center illustrate 10\% ellipticity.\label{fig4}}
\end{figure}

\begin{figure}
\plotone{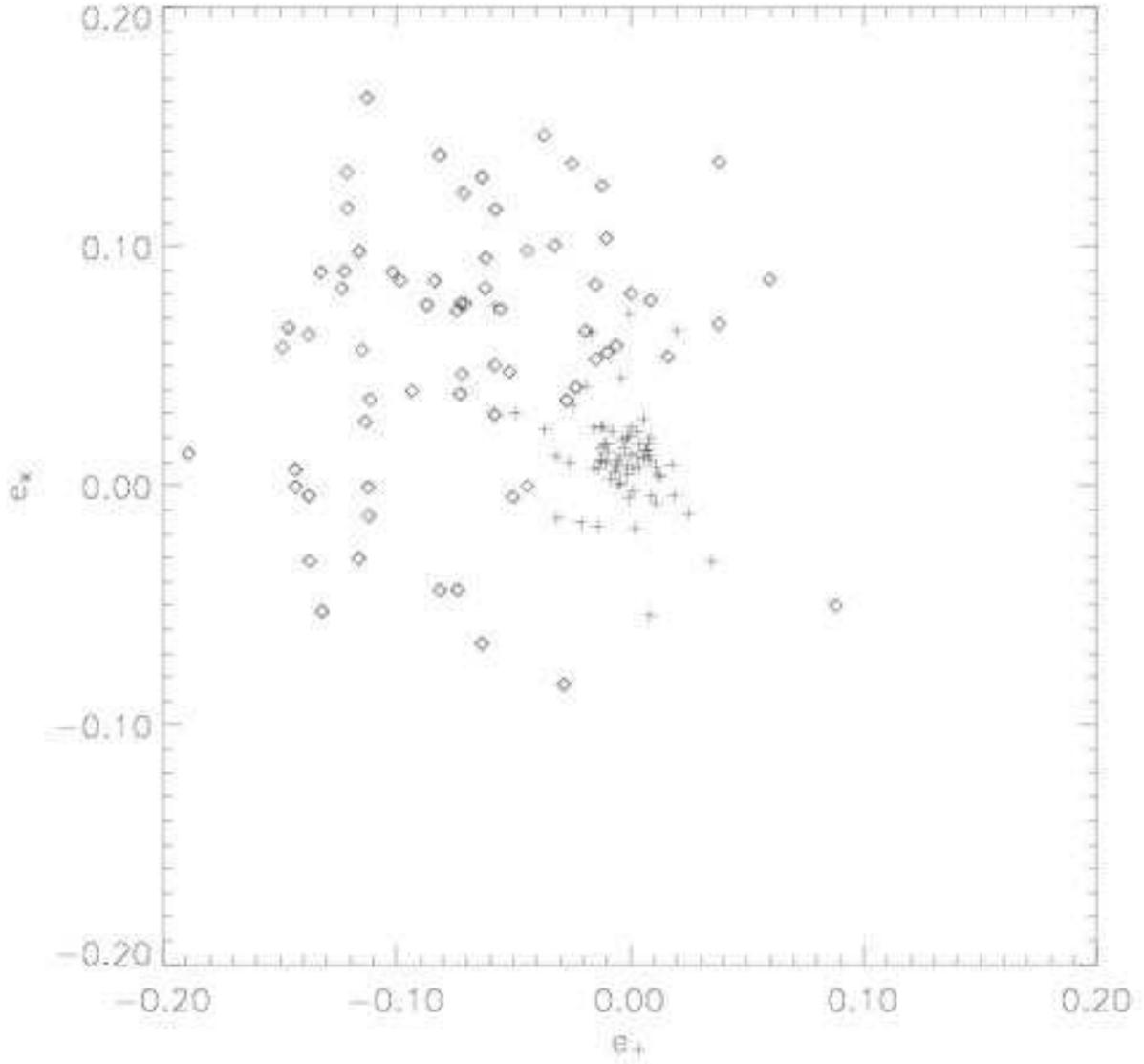}
\caption{PSF anisotropy removal on CL 0152-1357 field. Diamonds represent the initial ellipticities of stars
and plus symbols the corrected ellipticities after rounding kernel convolution. The improvement is remarkable
in both amplitude (size of scatter) and anisotropy (centroid). \label{fig5}}
\end{figure}

Using a typical half-light radius versus magnitude plot, we initially selected 73 isolated, bright stars which can
be used as local PSF indicators. We removed stars having any noticeable defects from the list by visual inspection, which
ended up a total of 62 stars. Figure~\ref{fig4}a shows the polarization pattern in CL 0152-1357 observation measured from these stars.
Then, using our model PSF we constructed ``rounding kernels'' \citep{ft97,kaiser00,bj02} which circularize the originally elongated PSFs and
applied them to the CL 0152-1357 images. Comparing the ellipticities before (Figure~\ref{fig4}a) and after (Figure~\ref{fig4}b) the application of the
rounding kernel verifies that our model PSF closely represents the real PSFs on the cluster image (see also 
Figure~\ref{fig5}). It appears that there still remain tiny but systematic residuals due to the incompleteness of the model; however, their effects on the cluster mass analysis are estimated to be negligible.

\subsection{PSF Correction from Deconvolution}

Though the ``rounded images'' obtained in the previous section can be used to measure the object shapes,
this is not preferred to the straightforward deconvolution technique
because of the following reasons. First, the rounding kernel always degrades the original image seeing because 
the kernel size must be comparable to the instrument PSF size in order to remove the anisotropy sufficiently, which
in particular is detrimental to very small galaxy images. Second, the dilution (circularization) correction provided by BJ02 is still
an approximation and \citet{hs03} showed in their simulation that indeed the prescription by BJ02
is not accurate if the kurtosis of the PSF is not small or the galaxy is not well-resolved. 

Due to the simple transformation rule of the Gaussian functions, the deconvolution can
be effected by convenient matrix manipulations:
\begin{equation}
 b^{o}_{p_o q_o} =  \sum C_{p_o q_o}^{p_i q_i p_s q_s} b^{i}_{p_i q_i}  b^{s}_{p_s q_s},
\end{equation}
\noindent
where $b^o$, $b^i$, and $b^s$ are shapelet components of the convolved image, the pre-seeing image, and the PSF, respectively.

The evaluation of the matrix elements $C_{p_o q_o} ^{p_i q_i p_s q_s}$ is summarized in BJ02 
(see also R03 for Cartesian coordinates).
After contracting $C_{p_o q_o} ^{p_i q_i p_s q_s}$ and $b  ^s _{p_s q_s}$, 
we get
\begin{equation}
b ^o_{p_o q_o} = \sum P_{p_o q_o}^{p_i q_i} b^i_{p_i q_i}.
\end{equation}
Now $P_{p_o q_o}^{p_i q_i}$ can be inverted to compute the deconvolved image $b^i_{p_i q_i}$
from the PSF-convolved original image $b^{o}_{p_o q_o}$. Because we desire to make the matrix $ P_{p_o q_o}^ {p_i q_i}$
invertible and also minimize the noise amplification which is typical in every deconvolution problem, 
the expansion of the PSF in shapelets must be truncated appropriately and the characteristic size of
the object should be large enough compared to the size of the PSF.

\section{LUMINOSITY ESTIMATION \label{lum}}

\begin{figure}
\plotone{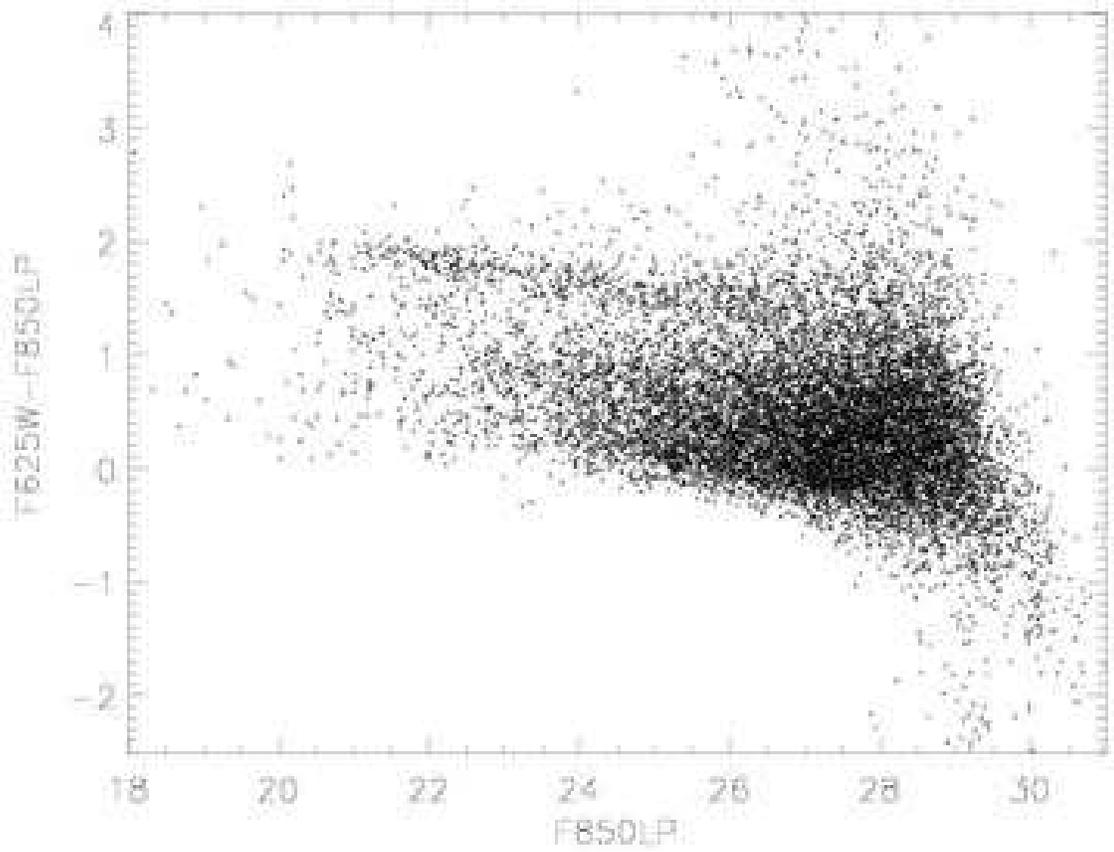}
\caption{Color-magnitude (CM) plot of CL 0152-1357. The tight CM relation of the early-type cluster galaxies
is present at $(r_{625}-z_{850})\sim1.9$. Bright stars are not removed.\label{fig5.2}}
\end{figure}

\begin{figure}
\plotone{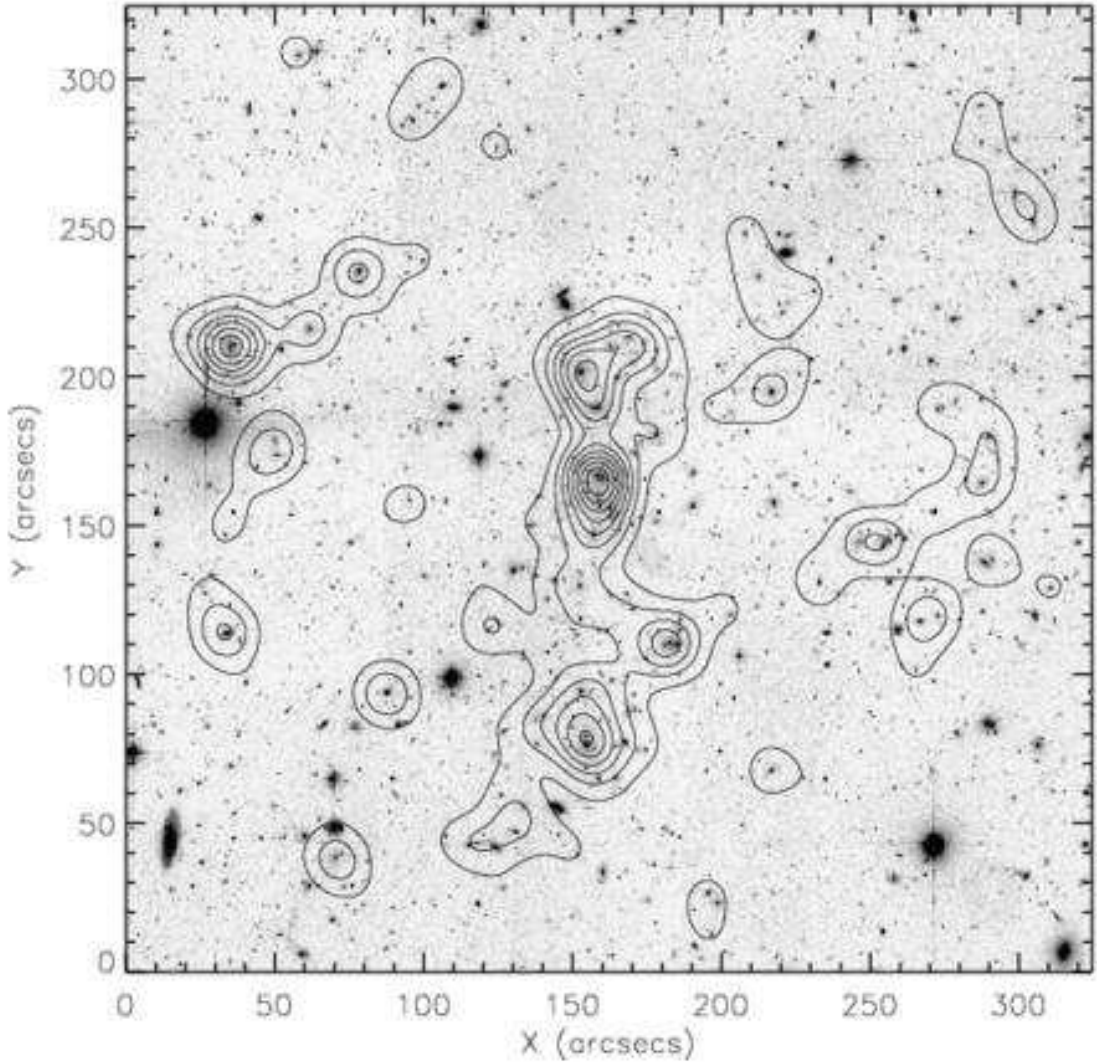}
\caption{Luminosity map of CL 0152-1357. For galaxies brighter than $z_{850}=22$, spectroscopically confirmed members are selected
whereas down to $z_{850}=25$ member selection is based on the color-magnitude relation of early-type galaxies. 
The luminosity center does not exactly lie on the BCGs, but is
slightly shifted to the south by $\sim2\arcsec$ because of the
presence of other bright cluster galaxies. 
\label{fig5.8}}
\end{figure}

We base our selection of cluster members on the tight color-magnitude (CM) relation 
of early-type galaxies of the cluster. Because the 4000\AA~break of the cluster ellipticals
is redshifted to the cutoff wavelength of $r_{625}$ filter, $(r_{625}-z_{850})$ colors are
better suited than $(i_{775}-z_{850})$ colors. As shown in Figure~\ref{fig5.2}, the bright cluster
red sequence of CL 0152-1357 occupies a relatively narrow strip in the $(r_{625}-z_{850})$ versus
$z_{850}$ CM diagram. Because increasing photometric errors at faint magnitudes cause
the distinction to become less apparent, we selected 371 galaxies brighter than $z_{850}\sim25$.
The spectroscopic catalog from VLT observations (R. Demarco et al. 2004, in preparation)
is used to reject bright non-cluster members ($z_{850} < 22$) and to include some known blue cluster
galaxies. We show the smoothed cluster light distribution from these member galaxies
in Figure~\ref{fig5.8}. The spectroscopic survey of the CL 0152-1357 field serendipitously discovered 
a foreground $(z\sim0.63)$ group of $\sim12$ galaxies rather loosely scattered over the entire field.
We excluded these galaxies in the above light distribution.
The vertically elongated main structure as well as the
less luminous but distinct clumps around the main body is clearly visible. We refer to the brightest concentration
in the light distribution as the $cluster$~$center$ hereafter. 
This smoothed light distribution
will be compared with those of the $Chandra$ X-ray and the weak lensing mass in \textsection\ref{substructure}.

In order to estimate the rest-frame luminosity of the cluster, we proceed as follows. 
From the dust maps of \citet{sfd98}, we obtained
E(B-V)=0.014 and determined the extinction corrections for $i_{775}$ and $z_{850}$ to be 0.028 and 0.020, respectively.
Then, synthetic photometry is performed by combining the latest ACS throughput curves (M. Sirianni et al. 2004, in preparation) and
the Kinney-Calzetti spectral templates \citep{kinney96} so as to establish the photometric transformation
of $i_{775}$ at $z\simeq0.84$ to the rest frame B magnitude. The linear best-fit result has the following form.
\begin{equation}
B_{rest} = i_{775} - (0.39 \pm 0.03) (i_{775}-z_{850}) + (0.76 \pm 0.05) - DM \label{photran},
\end{equation}
\noindent
where DM is the distance modulus of 43.63 in this cosmology and 
the uncertainties of the coefficients are estimated assuming an accuracy of $\sim$2\% in the synthetic photometry.
The total luminosty of the cluster is estimated by 

\begin{equation}
L_B = \sum_{j}  10 ^ { 0.4 (M_{B\sun} - B_{rest,j})} L_{B\sun},
\end{equation}
\noindent
where $M_{B\sun}=5.48$ is the absolute B magnitude of the sun. 

\begin{figure}
\plotone{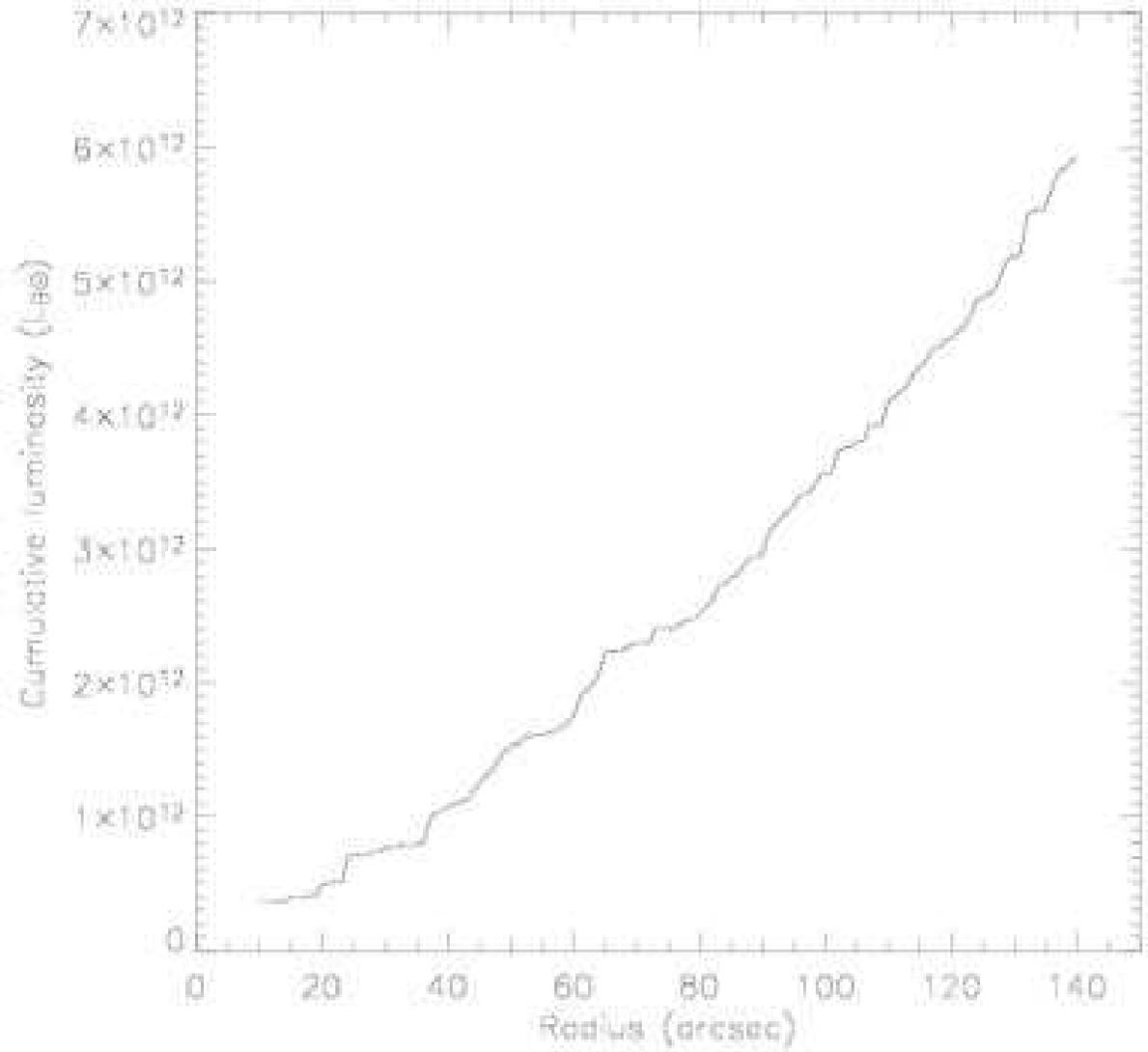}
\caption{Cumulative luminosity profile. The $i_{775}$ band magnitudes are transformed to the rest-frame B band absolute magnitudes after
corrections are made for the galactic dust extinction and the distance modulus. \label{fig5.7}}
\end{figure}

However, the $L_B$ obtained in this way does not include the contribution from
the faint ($z_{850} > 25$) population. In addition, we expect the blue cluster
members also comprise a significant fraction of the total light because
CL 0152-1357 is a high-redshift cluster. We choose to adopt the scheme by HFK00
in order to correct the total luminosity of the cluster for these factors. By fitting the
Schechter luminosity function to our sample galaxies, we found $\sim4$\% of the total light must be
added in order to account for the faint population. To estimate the fraction of the
blue cluster galaxies, we compared the spectroscopic catalog with
the CM selection and determined that we would lose $\sim20$ \% of the total luminosity if
not including the blue population. The fraction estimated here for the cluster CL 0152-1357 is slightly
higher than the value for MS 1054-03 obtained by HFK00 who quoted 16\%.
We present the cumulative light profile as
a function of the radius from the cluster center in Figure~\ref{fig5.7}. We observe that
the profile becomes marginally steeper as the radius approaches the field edges because of the increasing
contribution from blue cluster galaxies. The light profile, when reproduced without
including the spectroscopically confirmed blue population, showed no such trend.
In the WMAP cosmology, 1 Mpc corresponds to $\sim131\arcsec$ at $z=0.84$ and the total B-band luminosity
within this aperture becomes $5.2 \times 10^{12} L_{B\sun}$.

\section{MASS RECONSTRUCTION \label{massrecon_section}}  

\subsection{Source Galaxy Selection \label{source_galaxy}}
\begin{figure}
\plotone{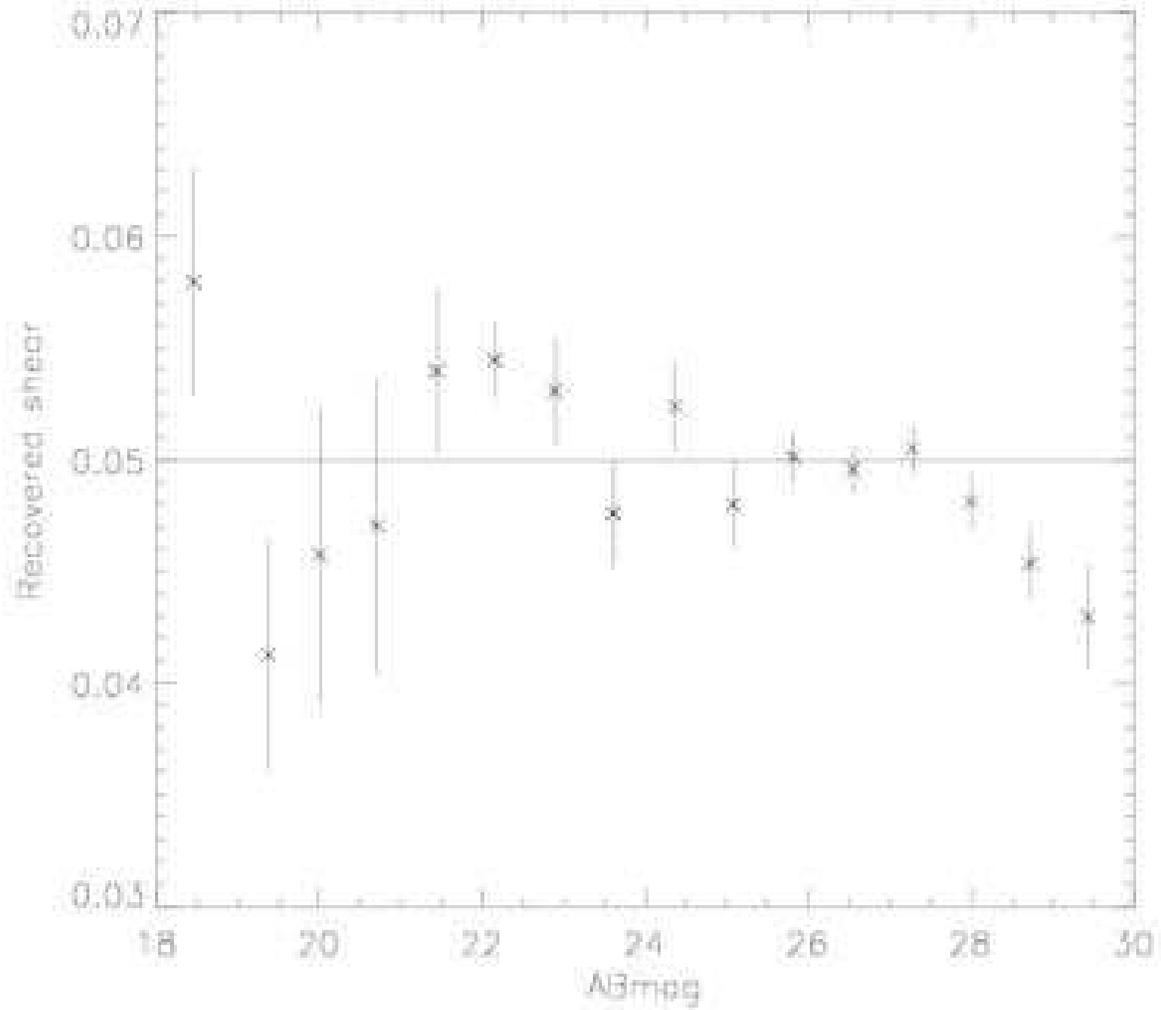}
\caption{Shear recovery test. The simulation is performed by artificially 
shearing the $r_{625}$ image of CL 0152-1357 by $\gamma = 0.05$ and measuring the difference in ellipticities of individual galaxies. The
large uncertainty for bright galaxies is due to the reduced number of objects. It is verified that down to $\sim$ 29.5 mag the
input shear is successfully recovered with less than $\delta \gamma \sim0.01$ deviation. \label{fig7}}
\end{figure}

\begin{figure}
\plotone{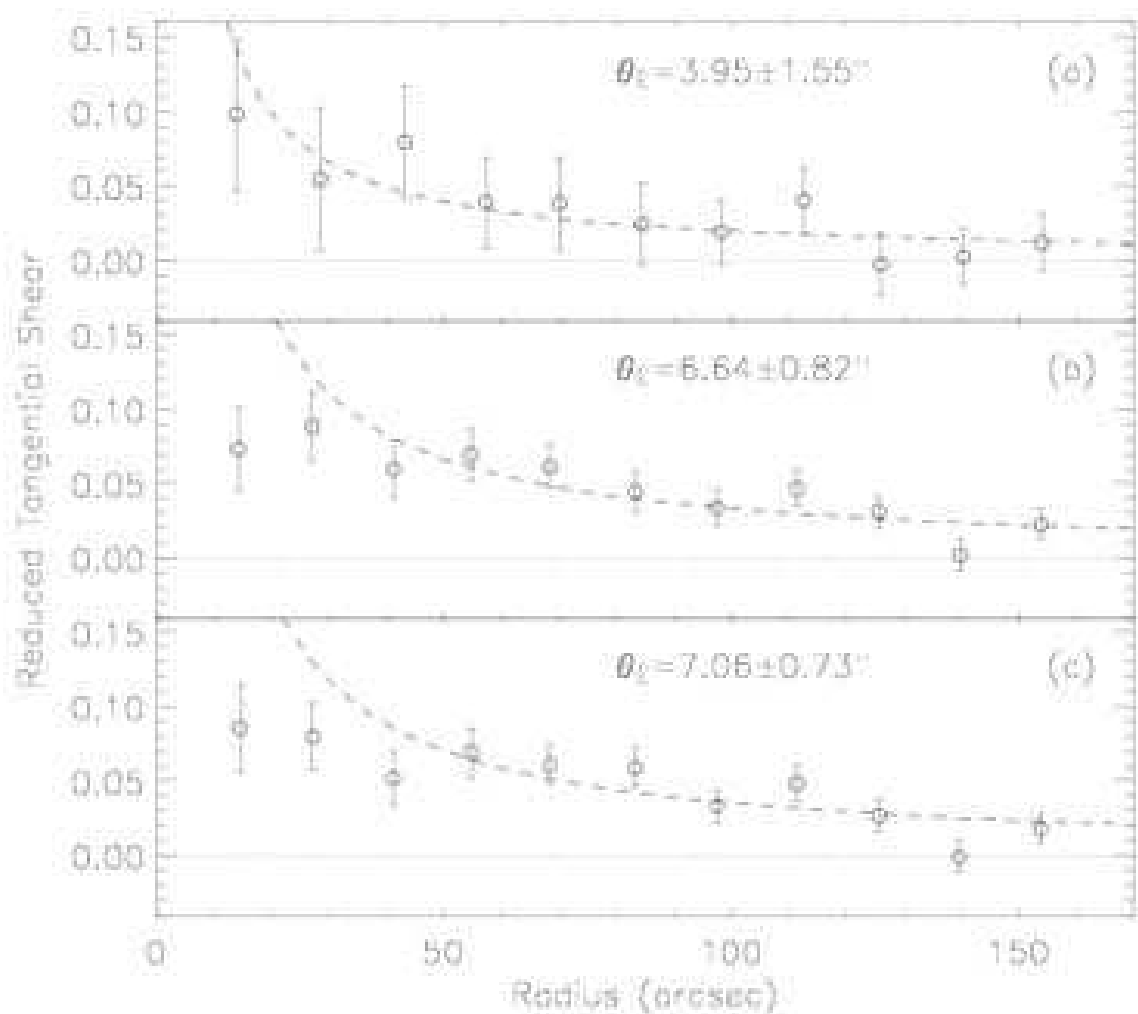}
\caption{Tangential shear versus radius from the cluster center. $\left < g_T \right >$ for
three different samples. The tangential shear from $bright$ sample ($z_{850}\le 26$; top)
is weak in comparison with those from $faint$ ($z_{850}\le 28.5$; middle) or $faintest$ ($z_{850}\le 30$; bottom) sample.
We also display best-fit singular isothermal sphere models (dashed) with corresponding Einstein Radius ($\theta_E$).
The inner region ($r \leq 50\arcsec$) is excluded from the fit.\label{fig6}}
\end{figure}

Careful selection of background galaxies must be made in both colors and
magnitudes in order to maximize the available signals. We rely on the
tight CM relation of the cluster to separate the early-type cluster members
from background galaxies. 
Galaxies whose $(r_{625}-z_{850})$ colors are bluer than $(-0.115z_{850} + 4.2)$ are
chosen. The distinction is not apparent at $z_{850} \gtrsim26$. In an idealized observation
where most of galaxies are well resolved, one can desire to include as faint galaxies as
possible because the distortion is greater for higher redshift objects. 
However, these faint
galaxies are in reality not only more susceptible to measurement noises, but also the amount of correction due to
the circularization of the PSF is greater, which increases the uncertainty of ellipticities by the same factor.
 To establish the
limiting magnitude for background galaxy selection, we carried out the following ``shear recovery''
test. After galaxy shapes are measured in the original field of CL 0152-1357, we artificially sheared
the entire image by 5 \% in real space. Then, the ellipticities of galaxies are determined once more
on the distorted image, and we checked how well the applied shear is recovered as a function of
magnitude (Figure \ref{fig7}). We observe that in spite of growing uncertainty as magnitude increases,
the shear is recovered down to $\sim$ 29.5 mag. Another useful experiment is to
examine the strength of the tangential shear while varying the magnitude limit of the sample.
Tangential shear is defined as
\begin{equation}
 \gamma_T  = -  \gamma_1 \cos 2\phi - \gamma_2 \sin 2\phi \label{tan_shear},
\end{equation}
\noindent
where $\phi$ is the position angle of the object with respect to the cluster center.
If no shear is present, the average of the tangential shear measured in the annulus around the
center must approach or oscillate near zero. However, if the shear is strong enough to
be measurable, $\left < \gamma_T \right >$ tends to be positive and the amplitude
is proportional to the magnitude of the shear.
We divided all the detected galaxies into ``bright'' ($24 \le z_{850} \le 26$), ``faint'' ($24 \le z_{850} \le 28.5$),
and ``faintest'' ($24 \le z_{850} \le 30$) samples, and the amplitude of azimuthal averages of the tangential shear
for different samples are compared (Figure~\ref{fig6}). Though it is not apparent whether or not
the signal from ``faintest'' galaxies is strongest, the amplitude of tangential shear from ``faint'' and ``faintest'' galaxies are undeniably 
greater than ``bright'' galaxies. 

We also examined the dependence of the lensing signal
on source galaxy colors by subdividing these ``faint'' and ``faintest'' samples into ``blue'' and ``red'' subsamples 
using $r_{625}-z_{850}$ colors.
We do not detect any significant change in shear strength between different color groups,
in contrast to the result of HFK00 who reported
their ``blue'' galaxies show stronger signals. However, the difference must be interpreted with
the different depth of the observations in mind. One plausible scenario is that 
the faint blue galaxies (FBGs) dominate the relatively low redshift, brighter background population whereas
the contribution from faint red galaxies becomes increasingly important for fainter background sources at high redshifts.

Considering the results of these experiments with the stability of deconvolution,
we choose ``faint'' galaxies as our ``best'' sample. The analysis hereafter is based 
solely on these galaxies.
The average number density of source galaxies 
in this sample reaches $\sim175 \mbox{arcmin}^{-2}$.   

\begin{figure}
\plotone{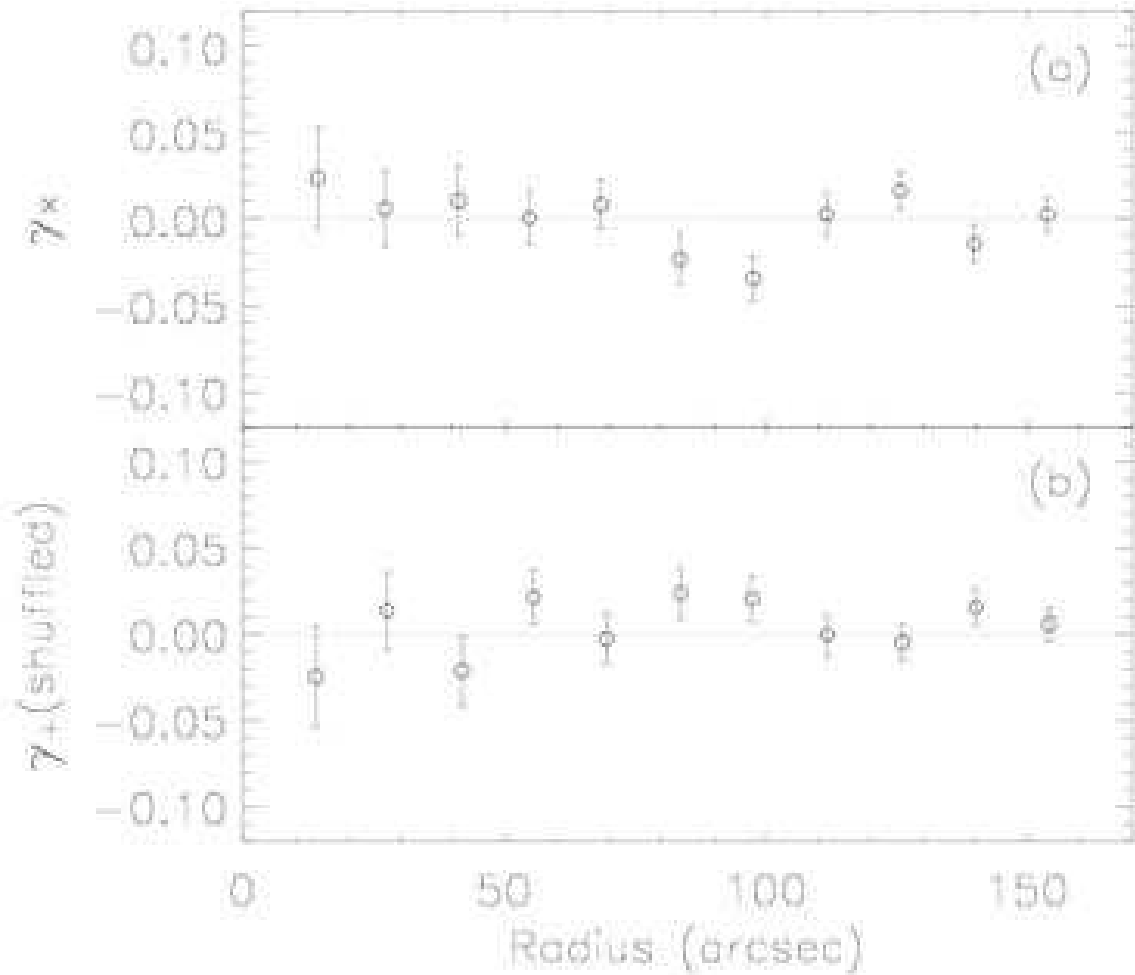}
\caption{Cross shear test for gravitational lensing.  Tangential shear is evaluated after
source galaxies are rotated by 45$\degr$ (top). The absence of the signal verifies that the
tangential shear in Figure~\ref{fig6} is due to the gravitational lensing. The comparison of this
result with the shuffle test (bottom) where galaxy ellipticities are randomly shuffled while 
the positions are fixed demonstrates that the cross shear scatters are consistent with the randomization test results.\label{fig6.5}}
\end{figure}

One of the useful methods to examine the significance of the lensing detection as well
as the systematics is to perform the following $null$ (cross shear) test. If the lensing signal
in Figure~\ref{fig6} is arising from the gravitational lensing, the resulting
shear must disappear when source galaxies are rotated by 45$\degr$ as shown
in the top panel of Figure~\ref{fig6.5}. We verify that the amplitude of the scatters are
consistent with that of the randomization test where source ellipticities are shuffled
while galaxy positions are held fixed. The bottom panel of Figure~\ref{fig6.5} shows the result from
one realization of this radomization.

\subsection{Redshift Distribution of Source Galaxies \label{redshift}}
Sufficient knowledge of the redshift distribution of source galaxies is essential in order
to achieve a proper scale in subsequent discussion of mass estimates. 
The critical surface density of the cluster is given by
\begin{equation}
\Sigma_c = \frac{c^2}{4 \pi G D_l \beta} 
\end{equation}

\begin{equation}
\beta = \left < \mbox{max} ( 0, \frac{D_{ls}} {D_s}) \right > \label{eqn_beta},
\end{equation}
\noindent
where $D_s$, $D_l$, and $D_{ls}$ are the angular diameter distance from the observer to the source,
from the observer to the lens and from the lens to the source, respectively.
Compared to low-redshift clusters, the critical surface mass density of high-redshift cluster is a relatively sensitive function
of source redshifts. That is, the angular diameter ratio $D_{ls} / {D_s}$ in equation~\ref{eqn_beta} for high-redshift clusters
changes more steeply than when the lens is at lower redshifts. 

\begin{figure}
\plotone{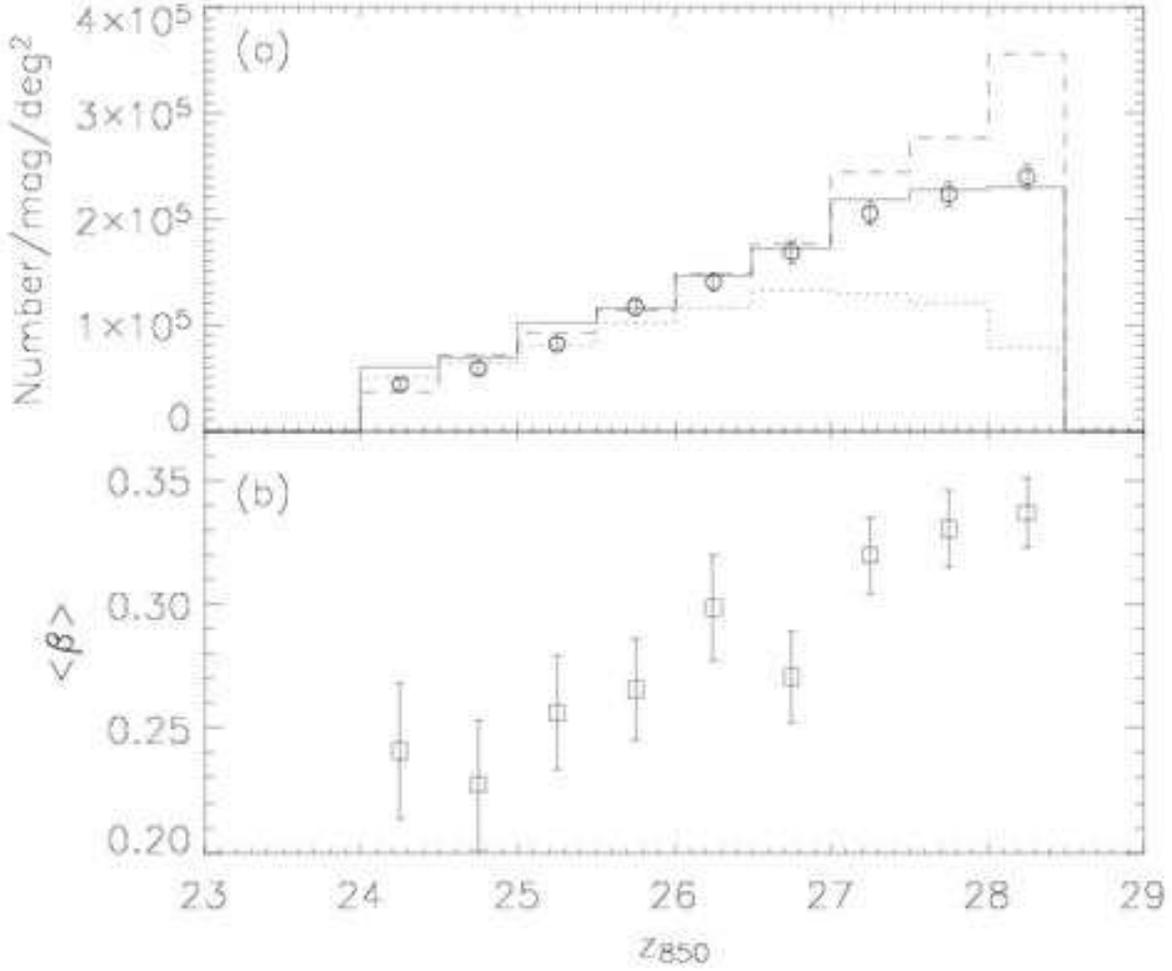}
\caption{(a) Contamination by cluster galaxies. The magnitude distribution of the source sample (solid) is compared
with those of the GOODS (dotted) and the UDF (dashed). Because $v_{606}$ filter is used in
both surveys instead of $r_{625}$, we transformed $(v_{606}-z_{850})$ color to $(r_{625}-z_{850})$ color 
(M. Sirianni et al. 2004, in preparation) to apply a consistent selection criterion. The fraction of
cluster galaxies in the source sample can be estimated from the magnitude distribution of the GOODS galaxies
down to $z_{850}\simeq26$. At $z_{850} \gtrsim26$, the incompleteness of the GOODS catalog is clear. In order
to estimate the contamination in this magnitude range, we degrade the UDF images to mimic the S/N ratio
of our cluster observation and detect source objects on these simulated images (open circle).
(b) Estimation of $\left <\beta \right >$ as a function of magnitude. The contamination by cluster members are not included
yet in this plot.
\label{fig_contamination}}
\end{figure}

It is certain that the selected source population in \textsection\ref{source_galaxy} actually contains
blue cluster members as well as foreground galaxies. The presence of these non-background galaxies
dilute the shear signal. We attempt to estimate the fraction of the cluster galaxy contamination in this sample
using the publicly available deep ACS images from the Great Observatories Origins Deep Survey \citep[GOODS;][]{giavalisco04} 
and the Ultra Deep Field \citep[UDF;][]{beckwith03}. The comparison of magnitude distribution of our source sample with
those from these two surveys enables us to determine the fraction of
cluster galaxies per magnitude bin up to the cosmic variance. Because F606W (hereafter $v_{606}$) passband is
used in both surveys instead of $r_{625}$, the $(v_{606}-z_{850})$ color is transformed to the $(r_{625}-z_{850})$ color 
to maintain the consistent selection criterion. At faint magnitudes ($z_{850} \gtrsim26$), the
GOODS catalog is incomplete. To infer the fraction of the cluster galaxies in this magnitude range, we add noise to 
the UDF images to match the SNR of our cluster observations and detect source objects on these degraded
images. It appears that the contamination from the cluster galaxies is not significant over the whole magnitude range (Figure~\ref{fig_contamination}a).

Because typical magnitudes of background galaxies are still far beyond
spectroscopic reach, we have to estimate the mean redshift of background galaxies via photometric redshift techniques. We
choose to use the photometric redshift catalog of the UDF (D. Coe et al. 2004, in preparation) to establish the
mean redshift of background galaxies in the cluster field. Combining the aforementioned ACS UDF with the NICMOS 
F110W and F160W (hereafter $j_{110}$ and $h_{160}$, respectively) observations,
we obtain reliable photometric redshifts of faint galaxies well beyond the faint end ($z\simeq28.5$) of our source
galaxy sample. The detailed description of the observation and the photometry
\footnote{We find that the photometric zeropoints of the NICMOS, $j_{110}^{AB}=23.4034$ and $h_{160}^{AB}=23.2146$,
released with the version 1.0 (2004 March 9) images still need to be refined. We adjust the
NICMOS zeropoints based on the spectral energy distribution (SED) fitting of the galaxies whose redshifts are
spectroscopically confirmed.} 
including the PSF matching across the different passbands
will be presented with the public release of the photometric redshift catalog. 
We generate the photometric redshift catalog of the UDF using
the Bayesian Photometric Redshift code \citep[hereafter BPZ]{benitez00}. The obvious advantage of the BPZ 
over the maximum-likelihood approach includes the use of the additional information on probability
distribution for given magnitudes termed $priors$. We take the redshift distribution of Hubble Deep Field North (HDF-N)
as priors, and the spectral template libraries of E, Sbc, Scd, and Im by \citet{cww80}
are selected. We also added two starburst templates (SB2 and SB3) by \citet{kinney96}. 
The recent post-launch throughput curves of ACS (M. Sirianni et al. 2004, in preparation) are incorporated into
the synthetic photometry of these SEDs.
The estimated $\left <\beta\right>$ as a function of $z_{850}$ is shown in
Figure~\ref{fig_contamination}b. In order to compute the final $\left <\beta \right >$ for the cluster, however, we need to
account for the relative number ratio of the background galaxies per magnitude bin as well as
the cluster galaxy contamination derived above. 
We assume that the foreground fraction of the sample is similar to what we obtain from the UDF.
Using the cosmological parameters considered in the current paper, we find $\left < \beta \right > = 0.282$.
This value corresponds to a single source plane at $ \left < z \right > \simeq1.30$.

The hypothesis that source galaxies are located in a single plane, though convenient, causes biases in
the measurement of the reduced shear $g=\gamma/(1-\kappa)$ because actual galaxies have a broad redshift distribution. The first
order correction is approximated by (Seitz \& Schneider 1997; HFK00)

\begin{equation}
\frac{g^\prime}{g} =  1 + \left ( \frac {\left < \beta^2 \right >}{\left < \beta \right > ^2} - 1 \right ) \kappa \label{gprime},
\end{equation}

\noindent
where $g^\prime$ is the uncorrected reduced shear from the single-source-plane assumption.
From the photometric redshift catalog and equation~\ref{gprime}, we determine that the reduced shear would be overestimated 
by $ (1 + 0.58 \kappa)$. We take into account this effect in the nonlinear mass reconstruction.

\subsection{Shear Estimation}

\begin{figure}
\plotone{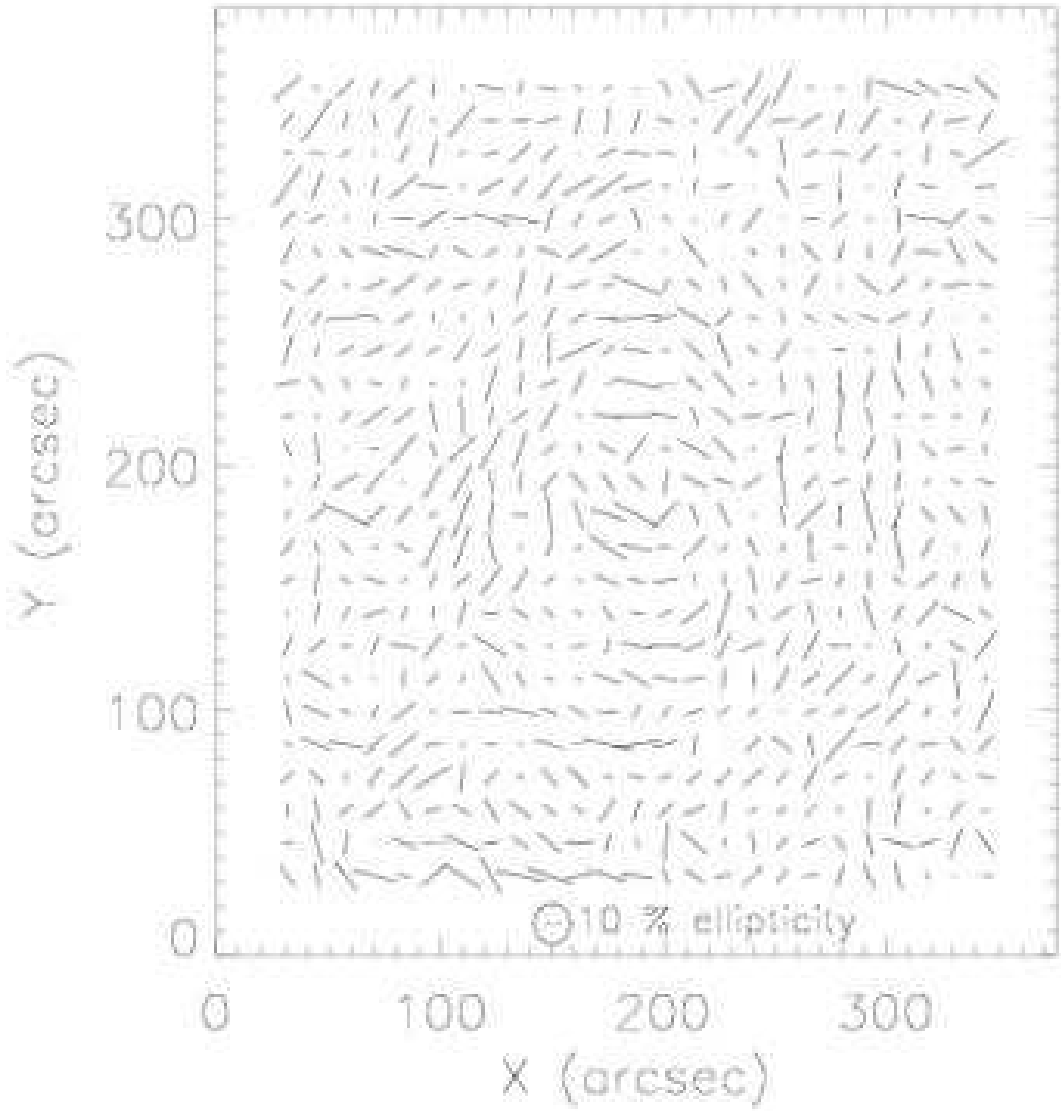}
\caption{Shear field of CL 0152-1357. Galaxy ellipticities measured from different passband images are
combined and smoothed with a Gaussian kernel (FWHM $\sim20\arcsec$).  \label{fig9}}
\end{figure}

Due to the intrinsic shapes of individual galaxies, the local shear at the given
location must be estimated statistically from a population of ellipticities. Naively taking a simple
arithmetic mean not only ignores the fact that the change in ellipticity $\delta e$ due to an applied shear $\gamma$
nonlinearly depends on the ellipticity of an object, but also discount the intrinsic ellipticity distribution
of source galaxies. HFK00 considered the weighting scheme combining measurement noise, intrinsic ellipticity, and
preseeing shear polarizability
in their weak lensing analysis of MS 1054-03. \citet{kaiser00} and BJ02 presented a similar, but more generalized
derivation incorporating the distribution of source ellipticities.
In this paper we adopted the ``easy'' weighting scheme suggested
in \textsection 5.2.3 of the BJ02 paper.
After smoothing the shapes of galaxies with the weighting function, we obtained the distortion field in Figure~\ref{fig9}.
The whisker plot obviously shows tangential alignments of background galaxies around the cluster center.
Since the distortion computed in this way is simply related to the shear by $\delta = 2 \gamma/(1+\gamma^2) \simeq2 \gamma$,
the shear map can be inverted to construct the parameter-free surface mass density of the cluster.

\subsection{Weak Lensing Mass Map}  \label{massrecon}

In the weak lensing regime ($\kappa \ll 1$), the shear $\gamma$ and the dimensionless mass density
$\kappa$ are simply related by the following convolution:

\begin{equation}
\gamma (\bvec{x}) = \frac{1}{\pi} \int D(\bvec{x}-\bvec{x}^\prime) \kappa (\bvec{x}^\prime) d^2 \bvec{x} \label{gamma} \label{gamma_of_k},
\end{equation}

\noindent
where $D(\bvec{x} ) = - 1/ (x_1 - i x_2 )^2$ is the convolution kernel.
By simple application of the convolution theorem, 
it is straightforward to invert equation~\ref{gamma_of_k} to express $\kappa$ in terms of $\gamma$ \citep[hereafter KS93]{ks93},

\begin{equation}
\kappa (\bvec{x}) = \frac{1}{\pi} \int D^*(\bvec{x}-\bvec{x}^\prime) \gamma (\bvec{x}^\prime) d^2 \bvec{x} \label{k_of_gamma}. 
\end{equation}

\begin{figure}
\plotone{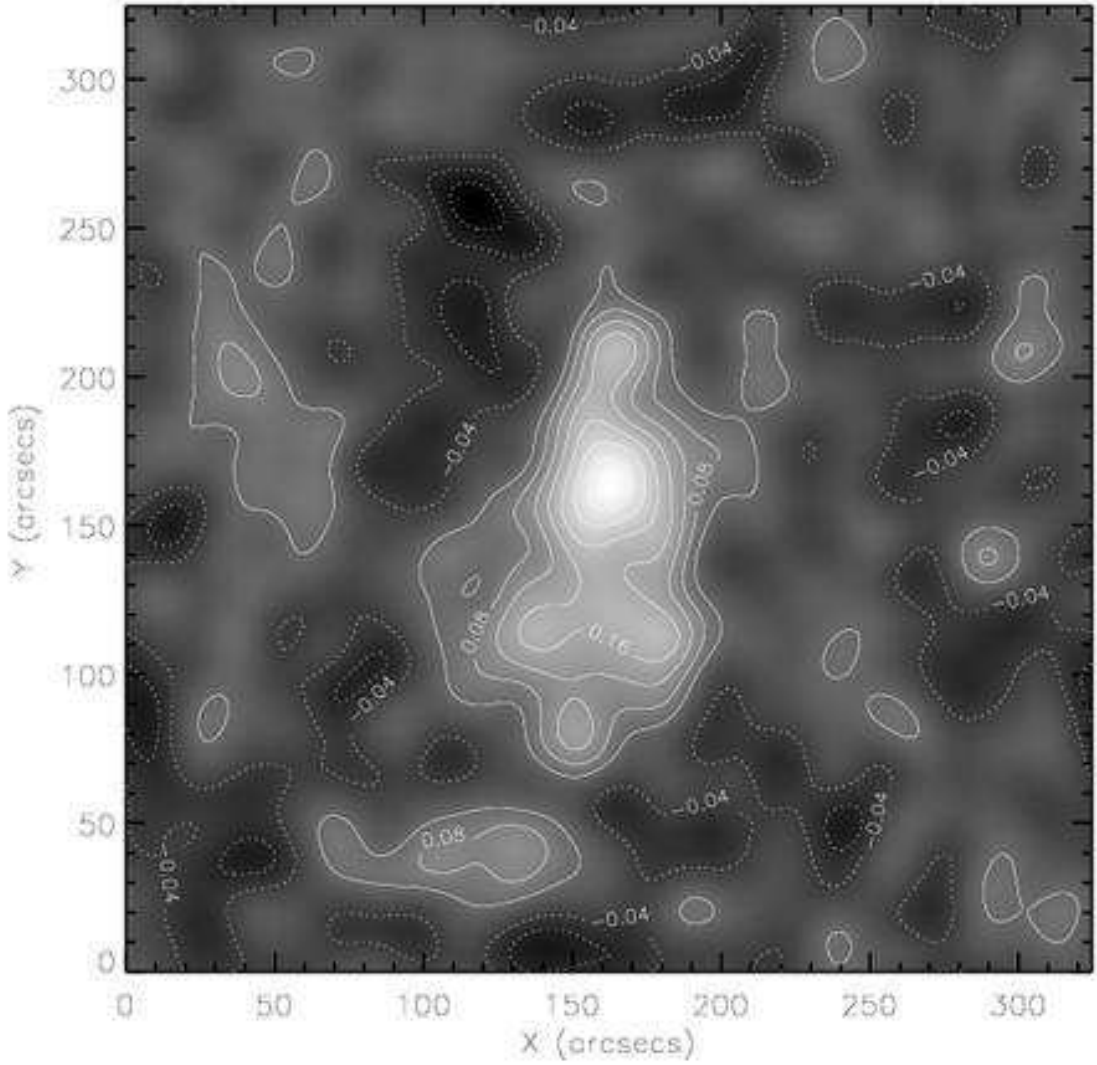}
\caption{The convergence $\kappa$ map reconstructed from the shear grid. The shear invariant transformation 
$\kappa \rightarrow \lambda \kappa + (1-\lambda)$ is not applied yet. The positions of distinct mass clumps are highly
correlated with the luminosity peaks found in Figure~\ref{fig5.8}.\label{fig9.5}}
\end{figure}

Therefore, in principle once $\gamma$ is reliably measured across the field, we can construct a parameter-free surface
mass density map either directly from equation~\ref{k_of_gamma} or indirectly through maximum-likelihood approach where
equation~\ref{gamma_of_k} or its variation is used as part of the likelihood function. The mass reconstruction 
obtained in this way is a useful tool to probe the substructure of the cluster. However, normally the direct computation 
of the physical mass from this map suffers from various artifacts of the reconstruction algorithms.
Most serious among them are spurious negative troughs around central mass clumps, thus causing the total
mass of the field to approach zero. This happens in part because typical variants of the original
KS93 mass inversion algorithms perform the convolution on a finite field. But, if the cluster
is assumed to be isolated, the boundary effect can be reduced greatly by simply extending the
field and extrapolating the measured shear outside the original field \citep{ss95a}.
The recent improvements such as maximum likelihood (probability) method \citep[e.g.,][]{bnss96,sk96,seitz98,mhgb02}
or unbiased finite-field inversion \citep[e.g.,][]{ss01,lb99}
can also minimize those artifacts without field extension. 
We experimented with three different algorithms: direct integration after smoothing \citep{ss95b}, maximum
likelihood \citep{bnss96}, and direct reconstruction based on variational principle
\citep{lb99}. We independently implemented the first two algorithms 
and the last one is kindly provided by the authors. We refer readers to the individual paper for details of each method.
We observe that the results from the above three algorithms
show no noticeable differences except for minor changes near the field boundaries where 
the signal is expected to be low and biased. We present the mass reconstruction
from the maximum-likelihood algorithm in Figure~\ref{fig9.5}. 

The resemblance of this mass reconstruction to the luminosity distribution (Figure~\ref{fig5.8}) is rather
remarkable. The vertically elongated central structure is clearly visible in the reconstructed mass map and the locations of dominant
mass peaks inside the main body agree well with those of cluster galaxy concentrations. We also note that even
outside the main body the spatial correlation between the mass overdensity and luminosity peaks is high.
We will discuss the detailed analysis and interpretation of the substructure in \textsection\ref{substructure}.

Despite the high-resolution of our mass reconstruction, the surface mass map in Figure~\ref{fig9.5} is not yet ready to
be used for inferring the physical mass unless the following ambiguities are resolved. First, what we measure is
the reduced shear $g=\gamma/(1-\kappa)$ rather than the true shear $\gamma$. That is, we always overestimate
$\gamma$ because of non-zero $\kappa$. The deviation cannot be ignored near the overdense regions where the assumption
$\kappa \ll 1$ breaks down. Besides, the correction due to the broad redshift distribution discussed in \textsection\ref{redshift}
becomes important if $\kappa$ is not sufficiently small. Nevertheless, it is possible to correct these effects by iteratively
updating $\kappa$ and $\gamma$ \citep {ss95a,ss97}. The initial mass reconstruction
can be carried out by setting $\kappa=0$. Now, since the improved information on $\kappa$ is available,
we can estimate the amount of correction for the original reduced shears $g$, and the updated $\gamma$ can be resubmitted into
the mass reconstruction. This $\kappa \rightarrow\gamma \rightarrow \kappa$ cycle converges typically after a few iterations.
The other more fundamental problem is that
the shear in equation~\ref{gamma_of_k} is invariant under $\kappa \rightarrow \lambda \kappa + (1-\lambda)$ transformation.
This so-called sheet mass degeneracy cannot be lifted unless external constraint is provided. Furthermore, without
the proper knowledge of this rescaling, we cannot achieve the previous nonlinear mass reconstruction 
because the validity of $\kappa$ for the use of updating shear $\gamma$ is not guaranteed.
In the next section we demonstrate that this rescaling of the surface density $\kappa$ map is in reality possible
with the help of the parameterized cluster profile. The parameterised mass modeling is safe from 
this sheet-mass degeneracy though the accuracy is sometimes compromised due to 
the inadequacy of the assumption of a particular mass profile.
We will also show that the direct mass estimates from this 
``rescaled'' $\kappa$ map are in good agreement with the results from the
conventionally favored aperture densitometry \citep{fahlman94,clowe98}.

\section{MASS ESTIMATES\label{massest}} 

In the current section we present the mass estimates of CL 0152-1357 through various routes. First, we
discuss the results from parameterized profile fitting. Especially 
for high-redshift clusters, which have a significant substructure, one does not expect that parameterized 
models can optimally describe 
complex profiles. Nevertheless, parameterized model fitting is an invaluable procedure not
only because it can easily provide a reasonable first-guess of the mass profile of the cluster, but also because
the results can be used to estimate the feasible mean surface mass density of the annulus far from the center,
thus enabling one to constrain $\bar{\kappa}$ of this region in subsequent mass estimation.
Second, we consider aperture mass densitometry which 
has been a preferred choice in
many cluster weak lensing studies because its implementation is straightforward and safe from artifacts arising
in most reconstruction algorithms. Finally, we attempt to measure
the mass of the cluster directly from the mass reconstruction map presented in the previous section. We show
that, after the consideration of proper rescaling $\kappa \rightarrow \lambda \kappa + (1-\lambda)$ and
non-linearity $g=\gamma/(1-\kappa)$, the mass estimate
obtained from the mass map is very close to the results from aperture densitometry.

\subsection{Parameterised Mass Profile}

\subsubsection{Singular Isothermal Sphere and Ellipsoid}
For a Singular Isothermal Sphere (SIS), the dimensionless surface density $\kappa$ is given as
\begin{equation}
\kappa = \frac{1}{2} \frac{\theta_E}{\theta} \label{eqn_kappa_sis},
\end{equation}
\noindent
where $\theta_E$ is the Einstein radius.
It is also easy to show that the simple relation $\left | \gamma \right | = \kappa$ exists for SIS.
After iteratively transforming the observed (reduced) tangential shears into the true
tangential shears $\gamma=(1-\kappa) g$, we fit a SIS profile to those tangential shears
in the annulus from 50$\arcsec$ to 160$\arcsec$.
From the typical $\chi^2$ minimization we find $\theta_E = 6.64\pm0.82\arcsec$.
In addition, we test if the ellipticity of the cluster can be detected by
fitting a Singular Isothermal Ellipsoid (SIE) \citep{ksb94}. While a single parameter can characterize SIS, SIE requires
3 parameters which describe Einstein radius $\theta_E$, axial ratio $f$, and orientation 
angle $\alpha$. Following the notation of \citet{ks01}, the surface mass density
and shear are related as follows:
\begin{equation}
\kappa (\theta,\phi) = \sqrt{\frac{f}{2}} \frac{\theta_E}{\theta} \left ( 1+f^2+(1-f^2) \cos \left (
2 (\phi-\alpha) \right ) \right) ^{- \frac{1}{2}},
\end{equation}
\noindent
where $f$ and $\alpha$ are axis ratio and orientation angle of a cluster, respectively. The corresponding
shears are 
\begin{equation}
\gamma_1=-\kappa \cos(2\phi)
\end{equation}
\begin{equation}
\gamma_2=-\kappa \sin(2\phi).
\end{equation}

For SIE profile fitting we use the smoothed distortion field in the same annulus as in SIS fitting.
The best-fit parameters are $\theta_E = 7.19\pm0.72\arcsec$, $f= 0.36\pm0.10$, and $\alpha=17.4\pm6.7^{\degr}$ (from the
vertical axis).
The orientation angle is consistent with the distribution of the cluster galaxies as well as the mass reconstruction. 
The axial ratio $f$ indicates that the overall mass distribution is highly elongated and
the azimuthal variation of the surface mass density is still detectable even at large radii. 
We determine the velocity dispersions from the estimated Einstein radii of SIS and SIE to be
$\sigma_v^{SIS}=903^{+54}_{-57}$ km $\mbox{s}^{-1}$ and $\sigma_v^{SIE}=940^{+46}_{-48}$ km $\mbox{s}^{-1}$, respectively.
These results are in good agreement with the direct measurements from the redshift survey of the cluster
(R. Demarco et al. 2004, in preparation).
They measured the velocity dispersion from their
spectroscopic data within $\sim50\arcsec$ aperture radii and obtained $\sigma_v=919\pm168$ km $\mbox{s}^{-1}$
and $\sigma_v=737\pm126$ km $\mbox{s}^{-1}$ in rest-frame for the northern and southern clumps, respectively.
It is also possible to compare the velocity dispersions with the cluster
temperature estimates from the X-ray and Sunyaev-Zeldovich analyses.
Despite the fact that the $\sigma-T$ relation of the galaxy clusters is in general very scattered
along the theoretical prediction $\sigma_v \propto T^{1/2}$ line, the estimates of
$5.9^{+4.4}_{2.1}$ keV \citep{ebeling00}, $6.5^{+1.7}_{-1.2}$ keV \citep{della00}, $8.5^{+2.0}_{-1.5}$ keV \citep{joy01},
or $5.5^{+0.9}_{-0.8}$ keV \citep{maughan03} are consistent with these velocity dispersions. For example, if we adopt
the empirical relation ($\sigma_v / \mbox{km~s}^{-1})=10^{2.57\pm0.13}(kT/\mbox{keV})^{0.59\pm0.14}$ 
\citep{wu98}, we get $T \simeq4.5_{-1.3}^{+3.1}$ keV for the SIS velocity dispersion.

\subsubsection{NFW Profile Fitting} \label{nfw_fitting}

The NFW density profile \citep{nfw97} is defined as:

\begin{equation}
\rho(r)= \frac{\delta_c \rho_c(z)}{(r/r_s)(1+r/r_s)^2},
\end{equation}
\noindent
where $\delta_c$ is the halo overdensity expressed in terms of the concentration parameter $c$ as
\begin{equation}
\delta_c = \frac{200}{3} \frac{c^3}{\ln(1+c)-c/(1+c)} ,
\end{equation}
\noindent
$\rho_c(z)$ is the critical density of the Universe at the redshift of the cluster, and
$r_s$ is the scale radius of the profile.
The relation between mean surface density and gravitational shear is much more
complicated in the NFW profile. The useful mathematical formalisms
for gravitational lensing are worked out by \citet{bartelmann96}, \citet{wb00}, and \citet{ks01}.

Though many parameters seem to be involved in the characterization of the NFW profile, only two parameters are
independent. From the similar $\chi^2$ minimization as in SIS and SIE fitting, we estimate the concentration parameter $c$ and the 
scale radius $r_s$ to be $3.7\pm0.5$ and $40\pm6\arcsec$ ($309\pm45$ kpc), respectively. 
A virial radius is defined
as a radius where the mean interior density drops to 200 times the critical
density of the Universe at the redshift of the cluster. From the simple relation $r_{200}= c r_s$, the virial
radius is estimated to be $r_{200}=150\pm30\arcsec$ ($1.14\pm0.23$ Mpc).

\subsection{Aperture Mass Densitometry} \label{aper_mass}
\begin{figure}
\plotone{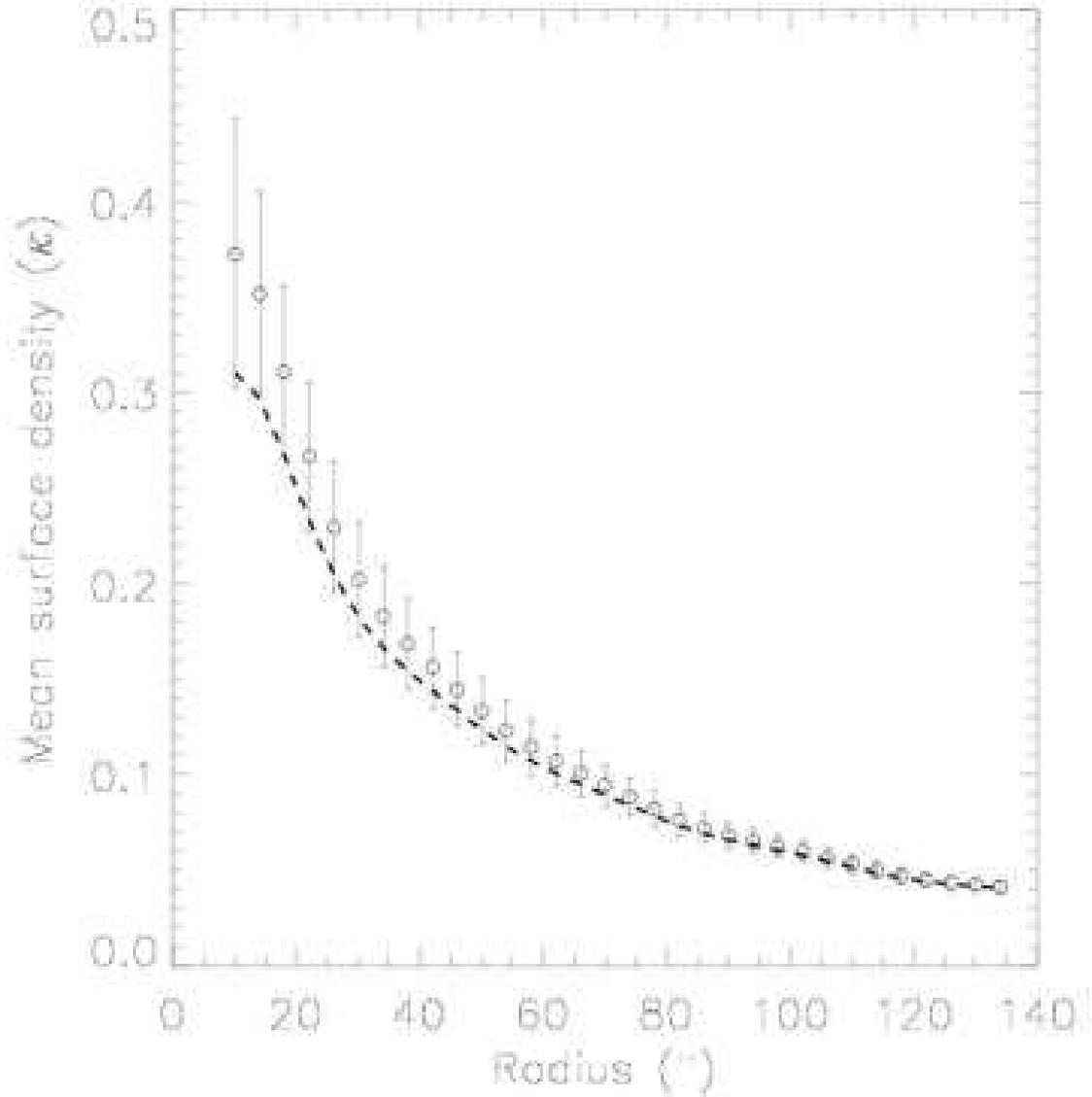}
\caption{Projected mean surface density within apertures in units of critical density $\Sigma_{cr}$. 
$\bar{\kappa} (140\arcsec < r < 160\arcsec) = 0.023 \pm 0.004$ is added to the $\zeta (r)$ statistic.
The dotted line represents the same mean surface
density when the $g = \gamma / (1 - \kappa)$ correction is included. This factor is important
at the cluster core where the $\kappa \ll 1$ assumption breaks down. \label{fig10}}
\end{figure}

When one's interest is to find a total mass within some given aperture radius $r$, the following
$\zeta (r)$ statistics provide a useful measure of lower limits on the mean surface mass density inside r:
\begin{equation}
\zeta _c (r_1,r_2,r_{max}) = 2 \int_{r_1} ^{r_2} \frac{
  \left < \gamma_T \right > }{r}d r    + \frac{2}{1-r_2^2/r_{max}^2} \int_{r_2}^{r_{max}}  \frac{ \left <\gamma_T \right >}{r} d r ,
\end{equation}
\noindent
where $\left < \gamma_T \right >$ is an average of tangential shears defined in equation~\ref{tan_shear}, $r_1$ is the aperture
radius, and $r_2$ and $r_{max}$ are the inner- and the outer radii of the annulus.
In the weak lensing regime where the relation between the measured ellipticity and the true shear
becomes linear, $\zeta _c (r_1,r_2,r_{max})$ can be directly computed from the
tangential shear to estimate $\bar{\kappa}(r<r_1) - \bar{\kappa}(r_2<r<r_{max})$
where $\bar{\kappa}(r_2<r<r_{max})$ is an average mass density in the annulus. In principle, if one choose $r_2$ and $r_{max}$
far enough from the cluster center to make the contribution vanishingly small, the $\zeta (r)$ above approaches
the genuine average surface mass density within the radius $r_1$.
In the present study, we used $r_2=140\arcsec$ and $r_{max}=160\arcsec$ 
in order to keep the entire annulus within the observed field. The mean surface density in this annulus
is expected to be low, but it still contributes to $\zeta (r)$. From the result of the SIS fit, we estimate
the dimensionless mean surface density of this region $\bar{\kappa}(r_2<r<r_{max}$) to be $0.023 \pm 0.003$.
Figure~\ref{fig10} shows the mean surface density inside given radii after the contribution
from the annulus $\bar{\kappa}(r_2<r<r_{max})$ is added. At small radii,
$\kappa$ is overestimated because of the reasons discussed in \textsection\ref{redshift} and \textsection\ref{massrecon}. 
The dashed line represents the mean surface density when this correction is applied.
The difference amounts to $\sim11$\% at $r\sim20\arcsec$.

\subsection{Rescaling of the Mass Reconstruction Map}
In \ref{massrecon} we discussed the general difficulties in translating the reconstructed convergence
map into the mass density in physical units. In order to lift the $\kappa \rightarrow \lambda \kappa + 1-\lambda$ 
degeneracy, we must be able to constrain $\kappa$ at least for a limited region of the field. This
is not impossible, however, because in \textsection\ref{aper_mass} we were able to estimate the mean surface density 
$\bar{\kappa}(r_2<r<r_{max})$ in the control annulus from the parameterized mass models. Therefore, we
can compare this with the value measured in the same annulus of the mass map. The
transformation parameter $\lambda$ becomes no longer arbitrary and can be determined from the relation
\begin{equation}
\lambda= \frac{\bar{\kappa}^{\prime}-1}{\bar{\kappa}-1},
\end{equation}
\noindent
where $\bar{\kappa}^{\prime}$ is a mean surface density of the annulus from the parameterised models and $\bar{\kappa}$
is the same quantity measured from the mass map. Then, we can apply $\lambda \kappa + 1-\lambda$ transformation
to the entire region of the mass map. Since this mass map is properly scaled, we can use it to
update the shear and feed this corrected shear back to the mass reconstruction. The procedure is iterated until convergence is reached.
In this study, no more than 4 iterations were needed. 

\begin{figure}
\plotone{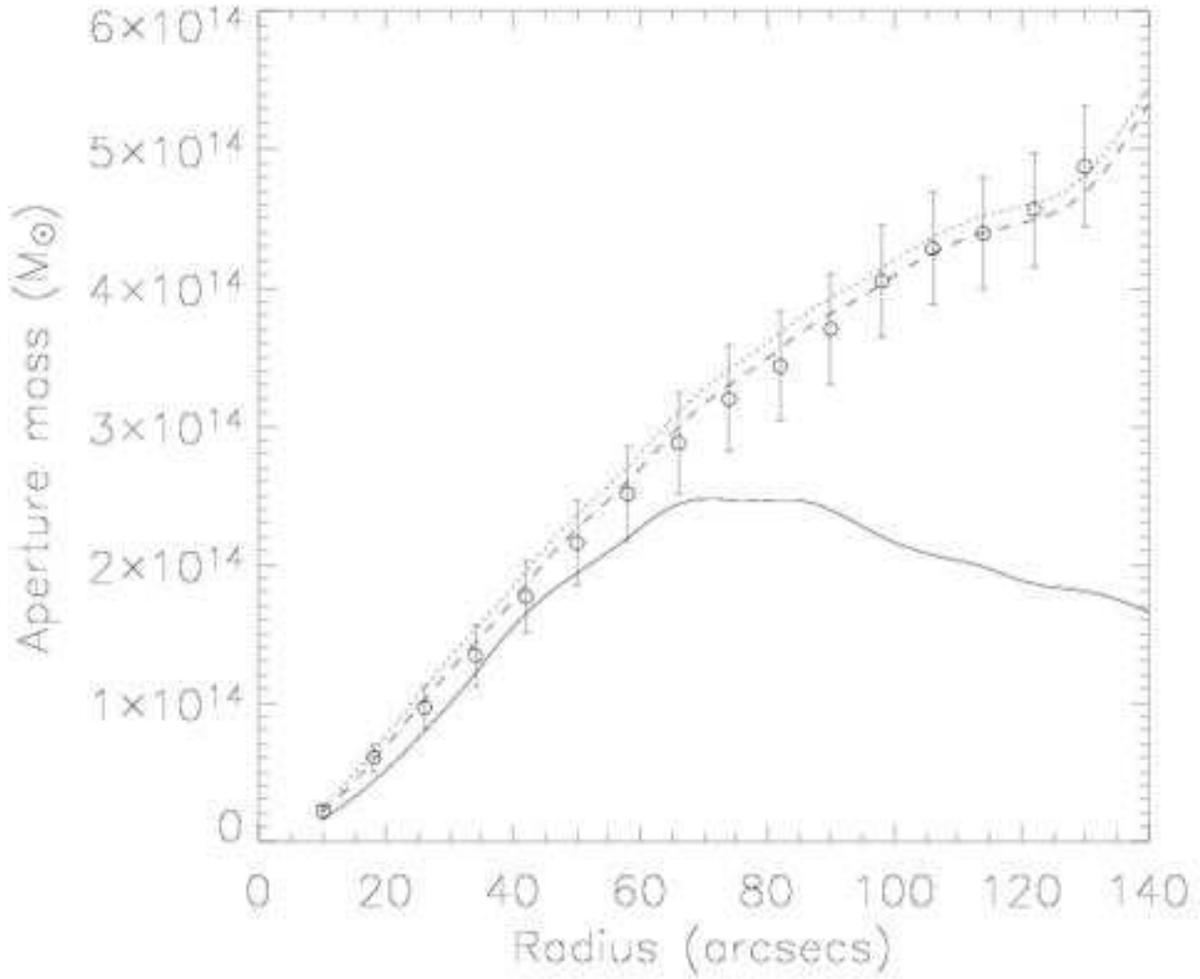}
\caption{Enclosed projected mass profile. The mass estimates derived from
the mass reconstruction are compared with those from the aperture densitometry
(open circles with error bars). The first, second, and final (4th) iterations of
rescaling of the mass map are represented by solid, dotted, and dashed lines, respectively.
 \label{fig11}}
\end{figure}

We compare the cumulative projected mass profile from this rescaled mass map
with the result from the aperture densitometry in Figure~\ref{fig11}. The excellent agreement between these two profiles 
demonstrates that at least in an azimuthally averaged sense
the mass estimate from the mass map is consistent with the result from the aperture densitometry.
In \textsection\ref{substructure}, we will also determine
the mass of the sub-clumps through both of these routes and show that the consistency can be generalized.
As far as the rescaling is appropriately calibrated, the use of the rescaled mass map
in probing the mass distribution of the cluster has obvious advantages over aperture densitometry.
The aperture densitometry always requires the control annulus to be set up
around target apertures, which is sometimes hindered if the annulus
cannot form a complete circle. Besides, to estimate the mean surface mass density inside the annulus,
one has to fit a particular parameterised mass model. If the discrepancy between
the assumed and the actual cluster mass profile is not small (e.g. due to the 
substructure), the procedure always introduces additional uncertainties. On the contrary,
the direct use of the rescaled mass map does not suffer from these obstacles, and this method can become
particularly useful when mass inside some arbitrary boundary needs to be estimated. 

\subsection{Comparison Between Parameterised and Parameter-Free Methods} \label{summaryofmass}

\begin{figure}
\plotone{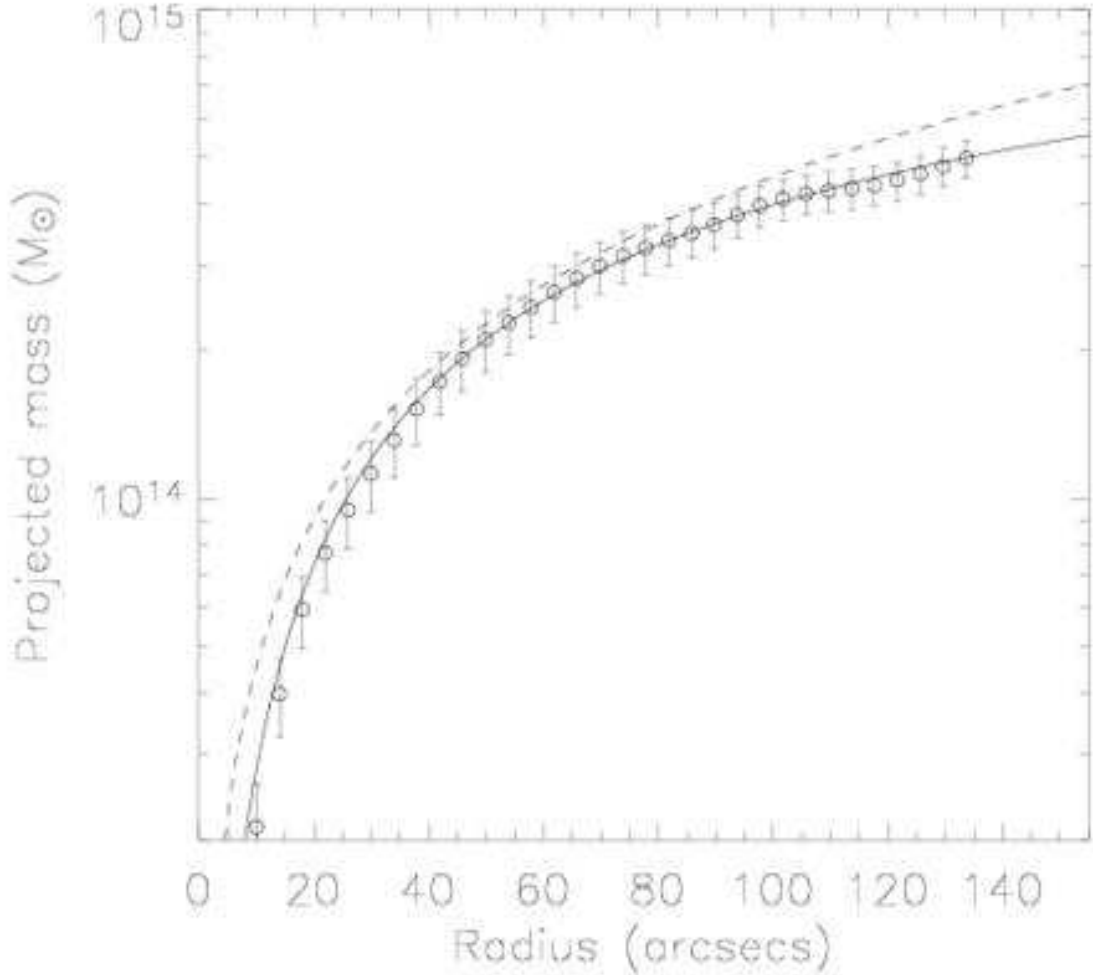}
\caption{Comparison of the parameter-free mass profile with the results from
SIS and NFW fitting. The parameter-free cluster mass profile (open circle with error bars)
is excellently described by the NFW profile with
a scale radius of $r_s =40\pm6 \arcsec$ ($309\pm45$ kpc) and a concentration parameter of $c=3.7\pm0.5$. The SIS
profile (dashed) overestimates the cluster mass by $\sim20$\% at $r\sim131\arcsec$ ($\sim$1 Mpc). \label{fig11.2}}
\end{figure}

The mass profiles from the SIS, NFW, and aperture densitometry are compared in Figure~\ref{fig11.2}.
We omit the SIE fitting result because it overlaps the SIS profile very closely.
Obviously, the actual cluster mass profile is best approximated by the NFW profile (solid). If the SIS (dotted)
is assumed instead, the total projected mass is overestimated by $\sim20$\% at $r\sim131\arcsec$ ($\sim$1 Mpc).
Considering the apparent filamentary substructure of CL 0152-1357 delineated by either
the light or the mass distribution, the excellent representation of
the azimuthal mass distribution of the cluster by the NFW profile is
rather remarkable. 
We summarize the mass estimates inside 1 Mpc radius
aperture in Table 1 for various combinations of cosmological parameters and methods.

\subsection{Mass-to-light Ratio} \label{section_ml}
\begin{figure}
\plotone{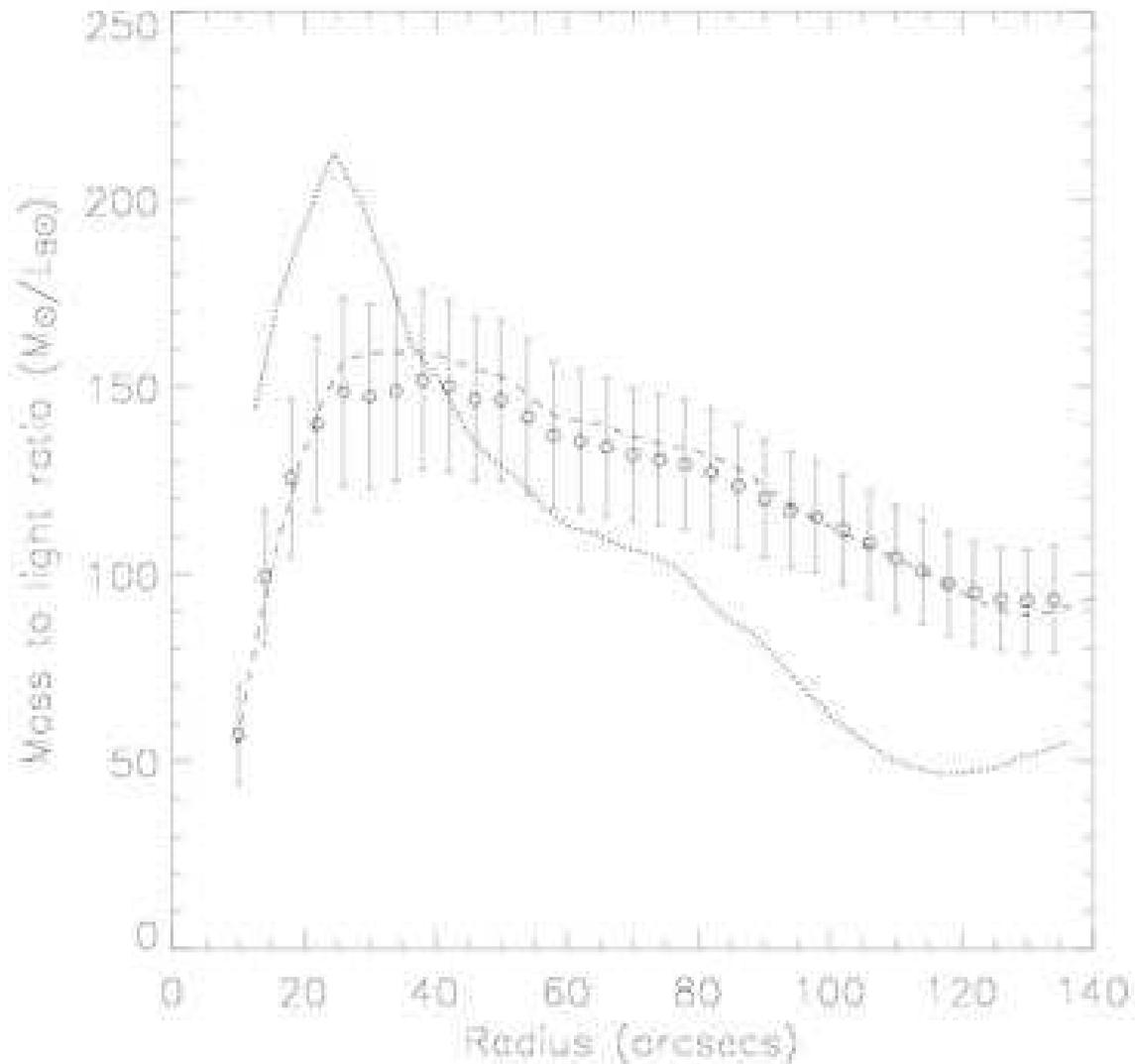}
\caption{Mass-to-light ratio profile of CL 0152-1357. Open circles and the dashed line represent the M/L ratio based on the
aperture mass densitometry and the reconstructed mass map, respectively. The dotted line shows the differential M/L ratio.
The uncertainties shown do not 
include the photometric errors for individual galaxies. The cumulative M/L ratio peaks at $\sim35\arcsec$ and then decreases
rather monotonically, reaching $\sim95 M_{\sun}/L_{B\sun}$ at 1 Mpc $(\sim131 \arcsec)$ from the cluster center. This
trend is more distinct in the differential M/L profile.
The light profile (Figure~\ref{fig5.7}) is smoothed to reduce the severe scatters caused
by the discrete galaxy positions. \label{fig12}}
\end{figure}

We present the mass-to-light ratio profile of CL 0152-1357
in Figure~\ref{fig12}. The cumulative mass-to-light ratio $M( \le r) / L_{B\sun} (\le r)$ (open circle and dashed) of the cluster rapidly rises to 
its maximum  
at $r\sim35\arcsec$ and then decreases rather monotonically. The decrease
of the profile looks more pronounced in the differential mass-to-light ratio $\delta M(r) / \delta L_{B\sun} (r)$ (dotted). 
It is verified that the profile when the blue cluster galaxies are excluded does not
significantly change though the slope is slightly reduced.
The small mass-to-light ratio near the cluster center seems to originate from the luminosity segregation of
the brightest cluster galaxies. \citet{cye97} studied the average mass-to-light profiles of 14 galaxy clusters from
the virial mass estimator. The resulting mass-to-light ratio averaged over substructure and asymmetries is 
high in the inner regions and gradually decreases until it starts to flatten at $r\sim0.7r_{200}$.
The average mass-to-light profile obtained from the kinematics and distribution of 3056 galaxies in 59 nearby clusters
in the ESO Nearby Abell Cluster Survey also shows a similar trend of a rapid rise followed by a gradual decrease
up to $r\sim0.7r_{200}$. Does the upturn of the M/L profile of CL 0152-1357 at $r\sim115\arcsec$ correspond to the beginning of the plateau
observed in those works? Assuming the feature is real and the empirical relation $r_{200}=r/0.7$ holds, 
the virial radius of the cluster can be evaluated to be $r_{200}\sim164\arcsec$
($\sim1.2$ Mpc). This value is surprisingly close to the independent estimation from NFW fitting in \textsection\ref{nfw_fitting}.

The average M/L ratio of the cluster within 1 Mpc aperture radius is estimated to be $95\pm8$ $M_{\sun} / L_{B\sun}$. It is
of interest to compare this value with the result for MS 1054-03 at a very similar redshift of $z\simeq0.83$.
In $\Omega_M=0.3$ and $\Omega_{\Lambda}=0.7$ cosmology HFK00 quoted the M/L ratio of $127 \pm10 h_{71} M_{\sun} / L_{B\sun}$, which
is higher than that of CL 0152-1357 by $\sim34$\%. The result is consistent with the still arguable, but popular
belief that the cluster M/L increases with richness. The M/L profile of MS 1054-03 by HFK00 shows
the gradual increase of the M/L ratio out to the field limit. If the M/L profile of MS 1054-03 is assumed to
conform to the aforementioned average M/L profile at large radii, we may suggest that the aperture
of 1 Mpc in MS 1054-03 field encompass only the inner region where the M/L is still high.

The logarithmic luminosity evolution, ln $(M/L_B) \propto (-1.06\pm 0.09)z$, is derived by \citet{vs03}
from massive cluster galaxies at $0.02\leq z \leq 1.27$. At $z=0.84$, the relation predicts
$\sim41$\% reduction in B-band luminosity and the M/L ratio of CL 0152-1357 is modified to
be $\sim232 M_{\sun} / L_{B\sun}$, which is similar to 
the results for other clusters \citep[e.g.,][]{cye97}. 

\section{Substructure of CL 0152-1357\label{substructure} }

\subsection{Mass Estimates of Individual Mass Clumps\label{bootstrap}}
\begin{figure}
\plotone{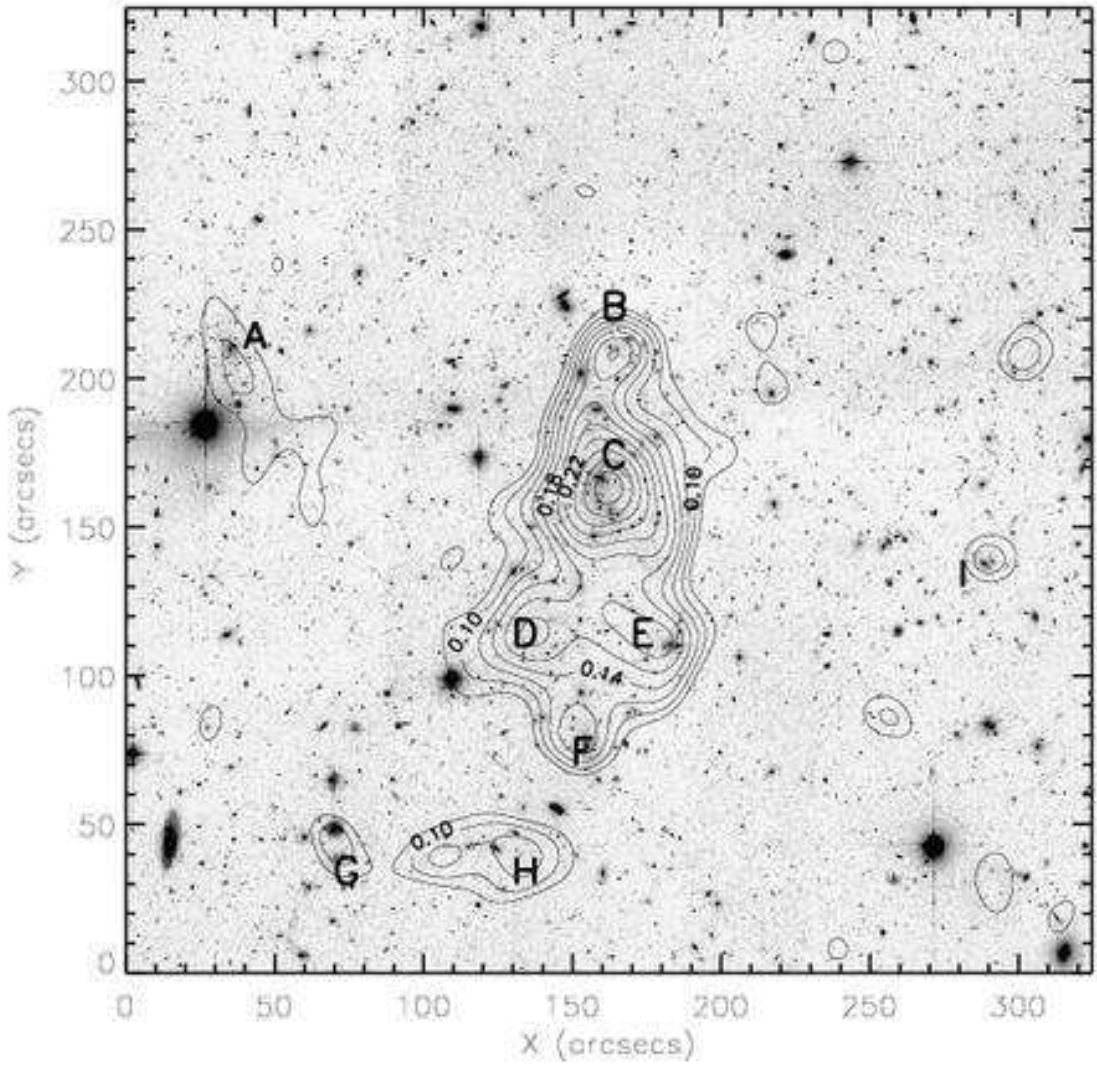}
\caption{Overlay of the mass map on the negative gray image. We only displayed contours for
$\kappa > 0.08$ which corresponds to $\sim3\sigma$. The mass estimates of annotated clumps 
are summarized in Table 2. \label{fig11.5}}
\end{figure}

\begin{figure}
\plotone{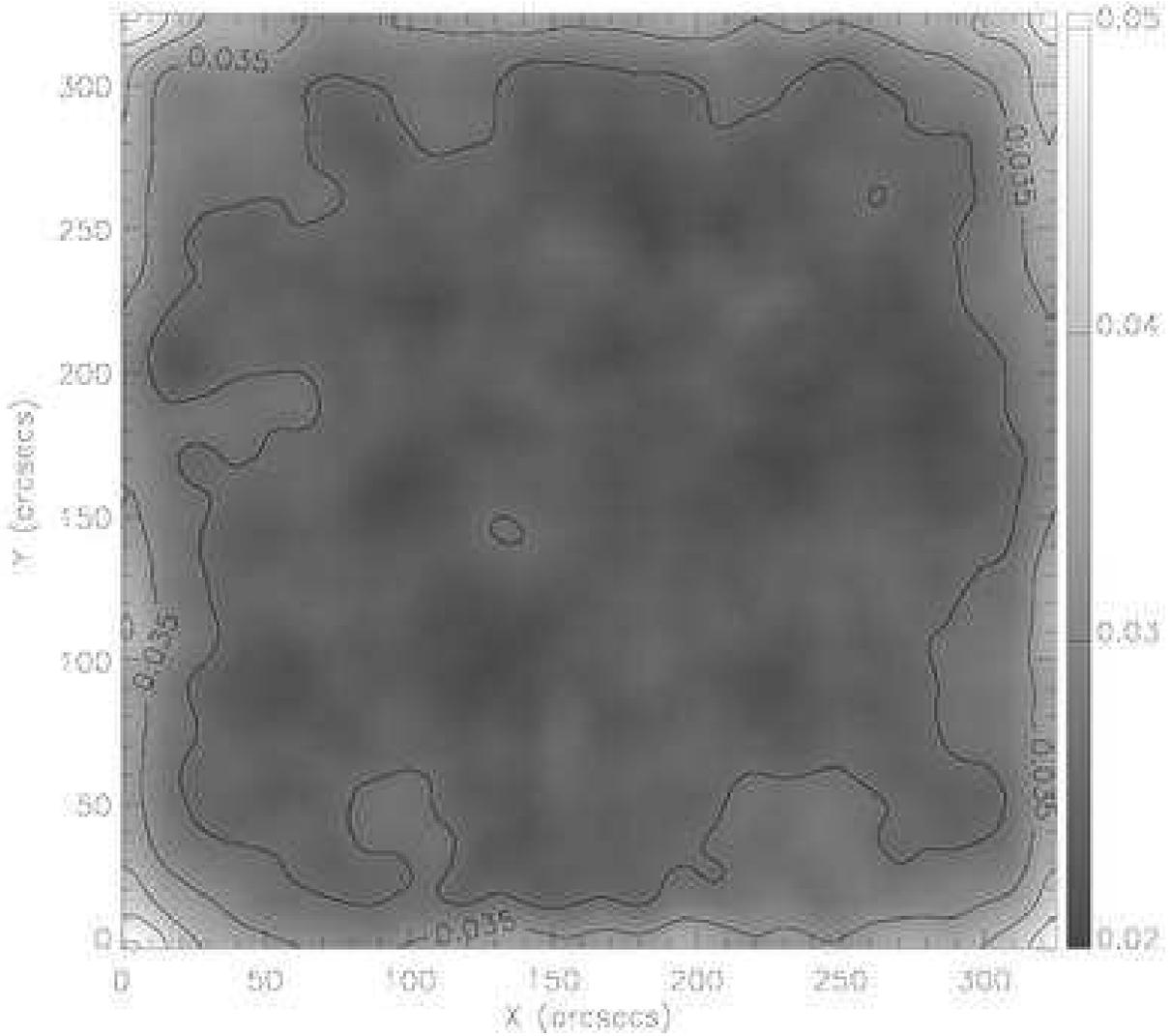}
\caption{RMS map of the rescaled mass map in Figure~\ref{fig11.5}. The rms 
is computed by 5000 runs of mass reconstruction. For each run, we randomly
resampled background galaxies from our source catalog. Note that at the field boundaries (especially
at the four corners) the uncertainties of mass pixels are higher than in the inner region. 
\label{fig_rmsmap}}
\end{figure}

Due to the high number density of background galaxies whose shapes are reliably measurable,
the reconstructed mass map reveals the cluster substructure in great detail. We overplot
the rescaled mass reconstruction on the negative gray image of CL 0152-1357 in Figure~\ref{fig11.5}.
We identified 9 mass clumps whose significance is above $\sim3\sigma$ and
galaxy counterparts are apparent. The significance for each mass pixel is computed by the use
of the RMS mass map (Figure~\ref{fig_rmsmap}), which is constructed from 
bootstrap 5000 realizations of mass reconstruction. These 5000 mass maps
are also used in \textsection\ref{otherstudies} to examine the uncertainties of the mass peak centroids.
The mass of these clumps within $20\arcsec$
aperture ($\sim150$ kpc) are computed
via the direct use of the reconstruction map as well as the examination of $\zeta(r)$ statistics (Table 2).
They are in good agreement with each other with overlapping uncertainties. 

\subsection{Comparison with Other Studies \label{otherstudies}}

\begin{figure}
\plotone{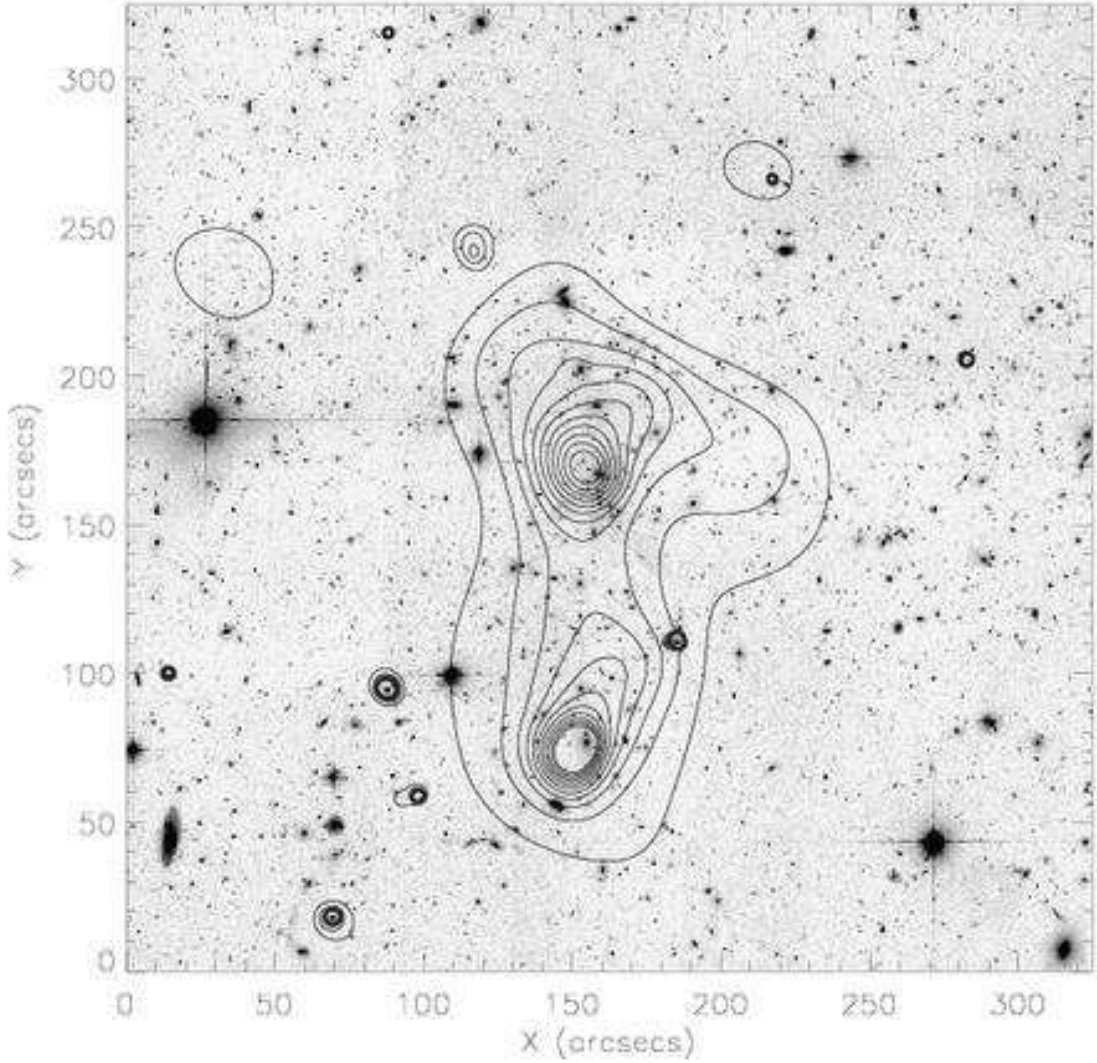}
\caption{Overlay of the smoothed X-ray map on the ACS detection image. The relative astrometric accuracy
between the $Chandra$ and the ACS images is $\sim1.4 \arcsec$ (see text). The X-ray flux map is generated
after adaptively smoothing the raw X-ray image using ``csmooth'' which is part of the Chandra Interactive Analysis 
of Observations Software (CIAO). 
\label{fig11.6}}
\end{figure}

Though the $Einstein$~$IPC$ first detected X-ray emission from CL 0152-1357,
its significance was not properly recognized because of the complex morphology of the emission.
The cluster was rediscovered in the WARPS \citep{scharf97}, 
the RDCS \citep{rosati98} and the SHARC \citep{romer00} survey. \citet{ebeling00}
analyzed the X-ray observation of CL 0152-1357 from the WARPS survey and showed that the X-ray 
morphology of the cluster is suggestive of very complex substructure that
can be also traced by cluster galaxy concentrations. The higher resolution of
$Chandra$ extended the work by \citet{ebeling00} and revealed two prominent X-ray
peaks \citep{maughan03}. In order to verify their results and also enable a direct comparison
of the X-ray morphology with our weak lensing mass distribution, we reanalyzed
the archival $Chandra$ observations. After adaptively smoothing the X-ray image via ``csmooth'', which is
part of the Chandra Interactive Analysis of Observations Software (CIAO), we obtained the
X-ray contour map of the cluster which we overlaid on the ACS image (Figure~\ref{fig11.6}). 
The astrometric accuracy of the ACS image with respect to the Guide Star Catalog 2 (GSC-2) is
$\sim0.05\arcsec$. Considering $\sim0.1\arcsec$ errors on the GSC-2 itself and also the absolute
astrometric accuracy of the $Chandra$ observation being $\sim1\arcsec$, we expect the alignment
between the two images are fairly precise, and the excellent agreements of the locations of X-ray point sources
with those of the optical counter parts confirm this fact. 
The two main diffuse X-ray peaks, though slightly off-center from two concentrations
of spectroscopically confirmed cluster members, seem to correspond to the mass clumps C and F in our
weak lensing map (Figure~\ref{fig11.5}). 
The mass clump A, which is associated with the $z=0.846\pm0.003$ group of galaxies (R. Demarco et al. 2004, in preparation),
is also detected as extended low surface brightness region in the $Chandra$ 
observation though this location is somewhat remote from that of the galaxy concentration.
In addition, one of the point-like X-ray sources appears to be associated with the mass clump E.
\citet{maughan03} examined the X-ray excess emission from the residual image
which is constructed by subtracting the best-fit model. They discovered that
the regions of excess emission lie midway between the two major clumps, stretched perpendicular
to the merging direction. They also found that the X-ray temperature of this region is
relatively higher. Though they were not conclusive because of the low significance of the feature,
the presence of a shock front was suggested.
It is remarkable that the feature
is also seen in the weak lensing mass map as clumps D and E. However, the cause of the
mass concentration in this region cannot be exclusively ascribed to a shock which originates
from the merger of two major subclusters. The cluster light distribution  
suggests that there may also be cluster galaxies associated with these clumps.

\begin{figure}
\plotone{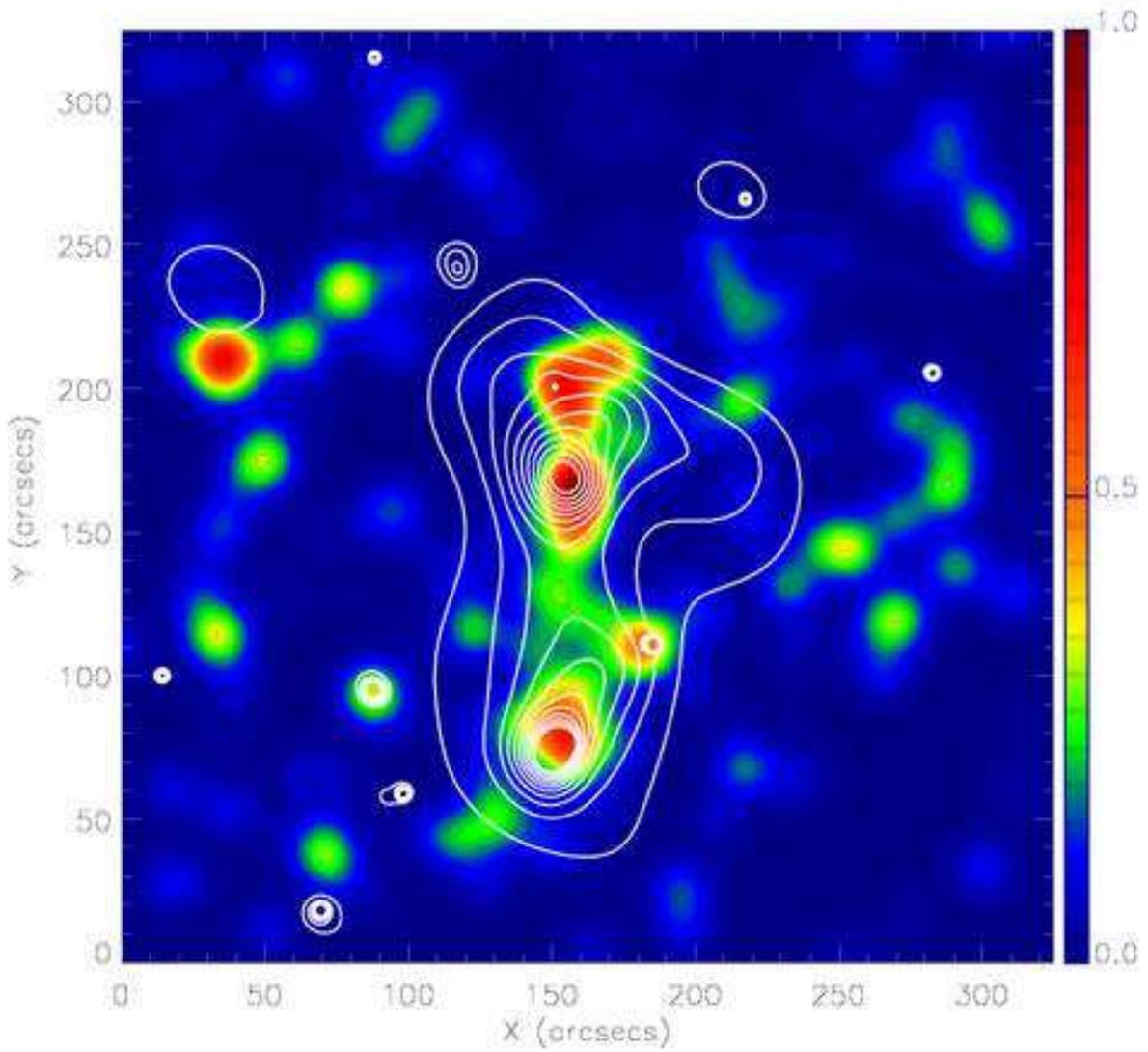}
\caption{Overlay of the smoothed X-ray contour (white solid) on the $i_{775}$ luminosity distribution (color-coded).
The X-ray peaks of two subclusters seem to be displaced away from the assumed merging direction.
\label{fig11.7}}
\end{figure}

\begin{figure}
\plotone{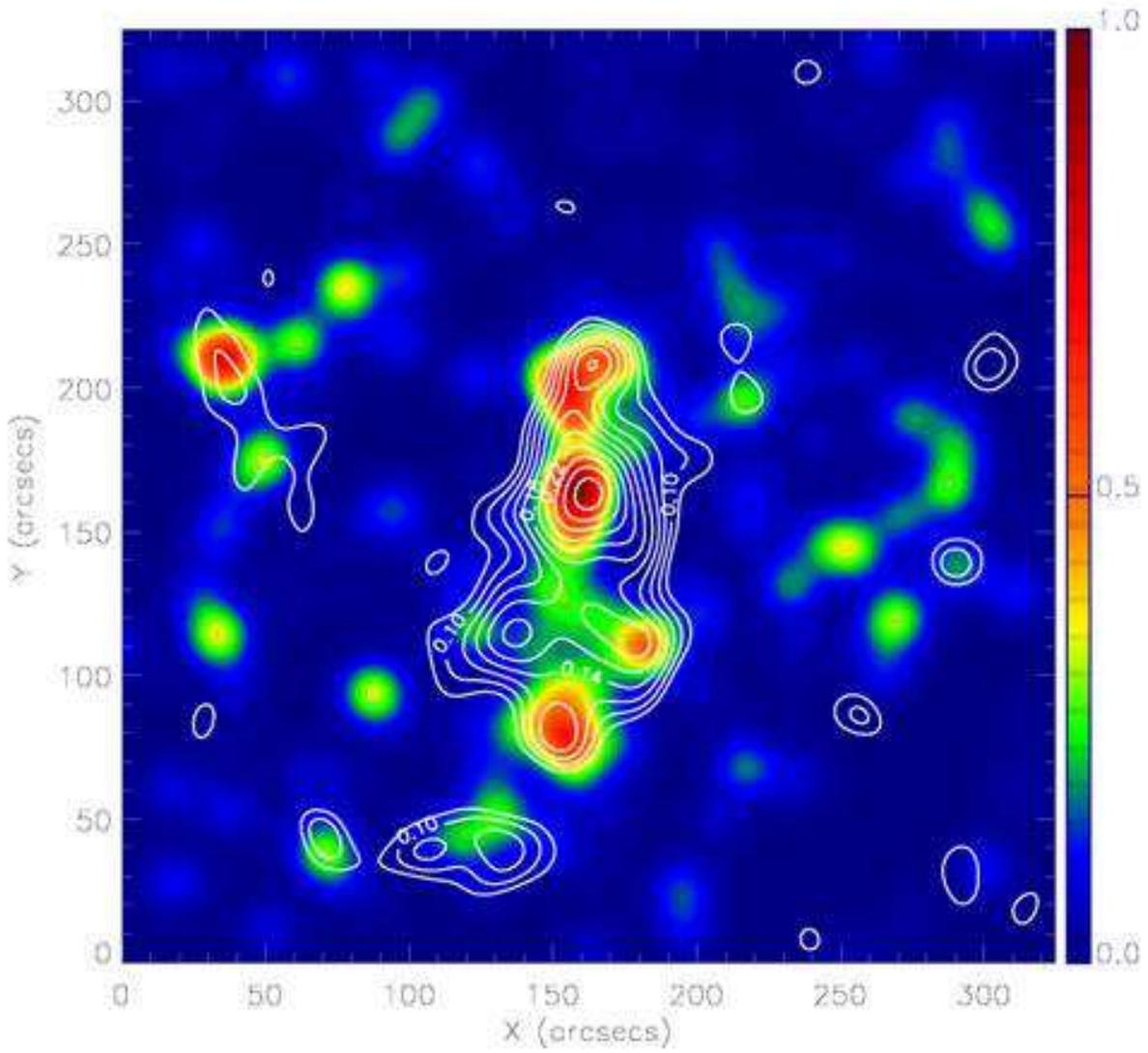}
\caption{Overlay of the mass contour (white solid) on the $i_{775}$ luminosity distribution (color-coded).  
The locations of mass clumps are in contrast to the X-ray peaks in Figure~\ref{fig11.7}, shifted toward the cluster center.
\label{fig11.8}}
\end{figure}

\begin{figure}
\plotone{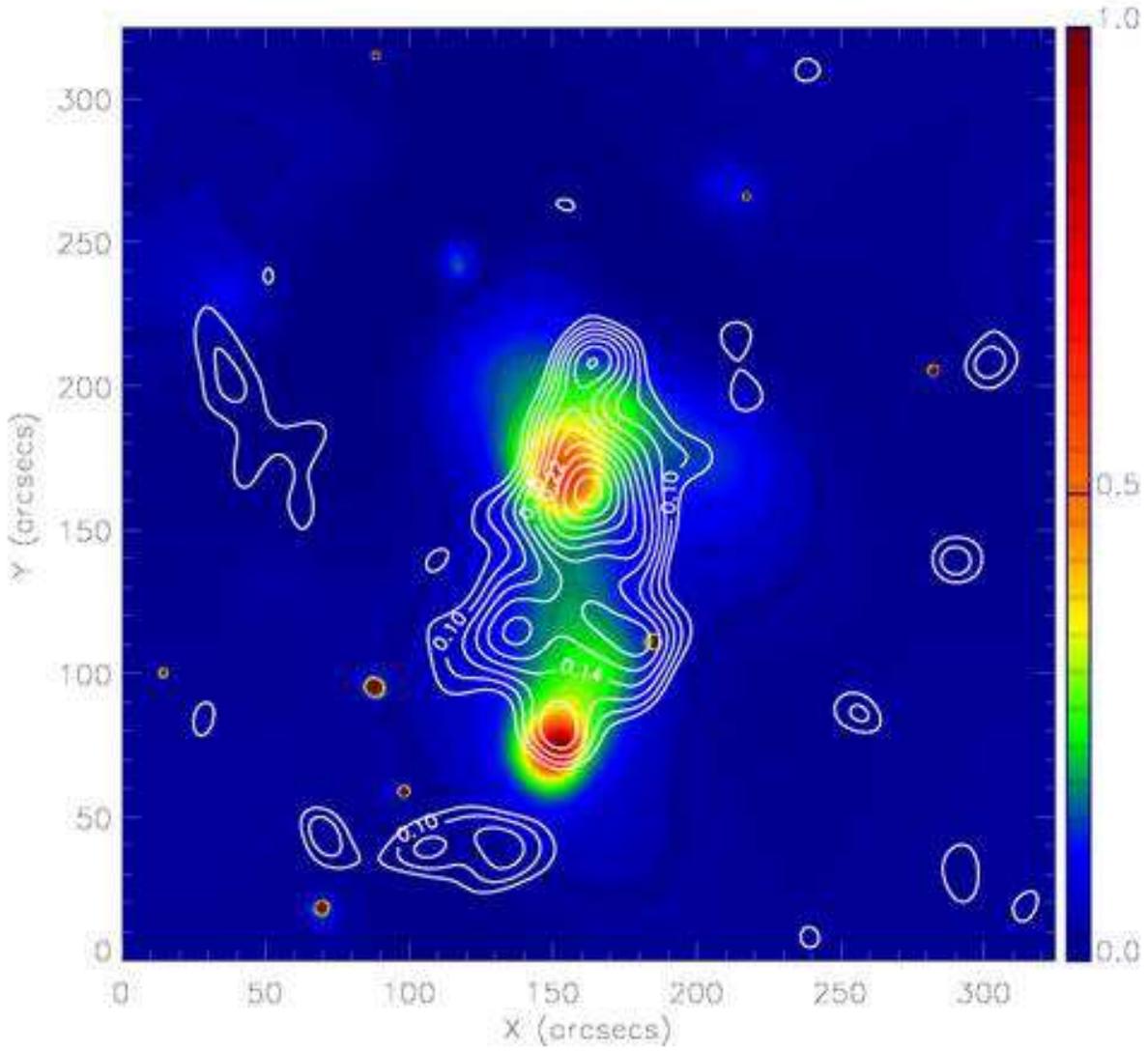}
\caption{Overlay of the weak lensing mass map (white solid) on the smoothed x-ray background
(color-coded). The offsets between mass clumps and X-ray peaks are distinct.
\label{fig11.9}}
\end{figure}

It is interesting to examine the displacements of peaks between the weak lensing mass map
and the X-ray flux contours. We present overplots for three different combinations:
 X-ray/optical (Figure~\ref{fig11.7}), mass/optical (Figure~\ref{fig11.8}), and mass/X-ray (Figure~\ref{fig11.9}).
In general, weak lensing mass distribution better traces the light distribution of the cluster as far as
the overall morphology and coincidence of clump locations are concerned. 
As \citet{maughan03} observed,
the location of the southern X-ray peak is offset by $\sim5\arcsec$ from the galaxies and
displaced away from the direction of the merger. They proposed that the displacement
is due to the fact that relatively collisionless galaxies are moving ahead of the viscous ICM
whose distribution is delineated by the X-rays.
We observe that the similar trend for the northern X-ray peak is also noticeable but with
smaller displacement. The offset is more obvious when the location of the northern X-ray peak
is compared with the smoothed luminosity center (not with the brightest galaxies; Figure~\ref{fig11.7}).
By examining the significance of the event counts around the northern X-ray peak, we verified
that the offset is not likely to be caused by artifacts of adaptive smoothing.
Similar, but more pronounced centroid offsets were 
detected in the $Chandra$ study of the merging cluster 1E 0657-56 \citep{markevitch02}. The X-ray
image shows that the X-ray centroids of two clumps are conspicuously displaced from the cluster
galaxy concentrations, suggesting the ram-pressure stripping of the gas components of the cluster.
If the dark matter is indeed collionless and is the dominant contributor to the total cluster mass, the location
of the mass peaks must be separated from the X-ray centroids. \citet{cgm04} performed
a weak lensing analysis of the same cluster and showed that the reconstructed mass peaks are
displaced from the X-ray halos while in good spatial agreements with the galaxy concentrations.

\begin{figure}
\plottwo{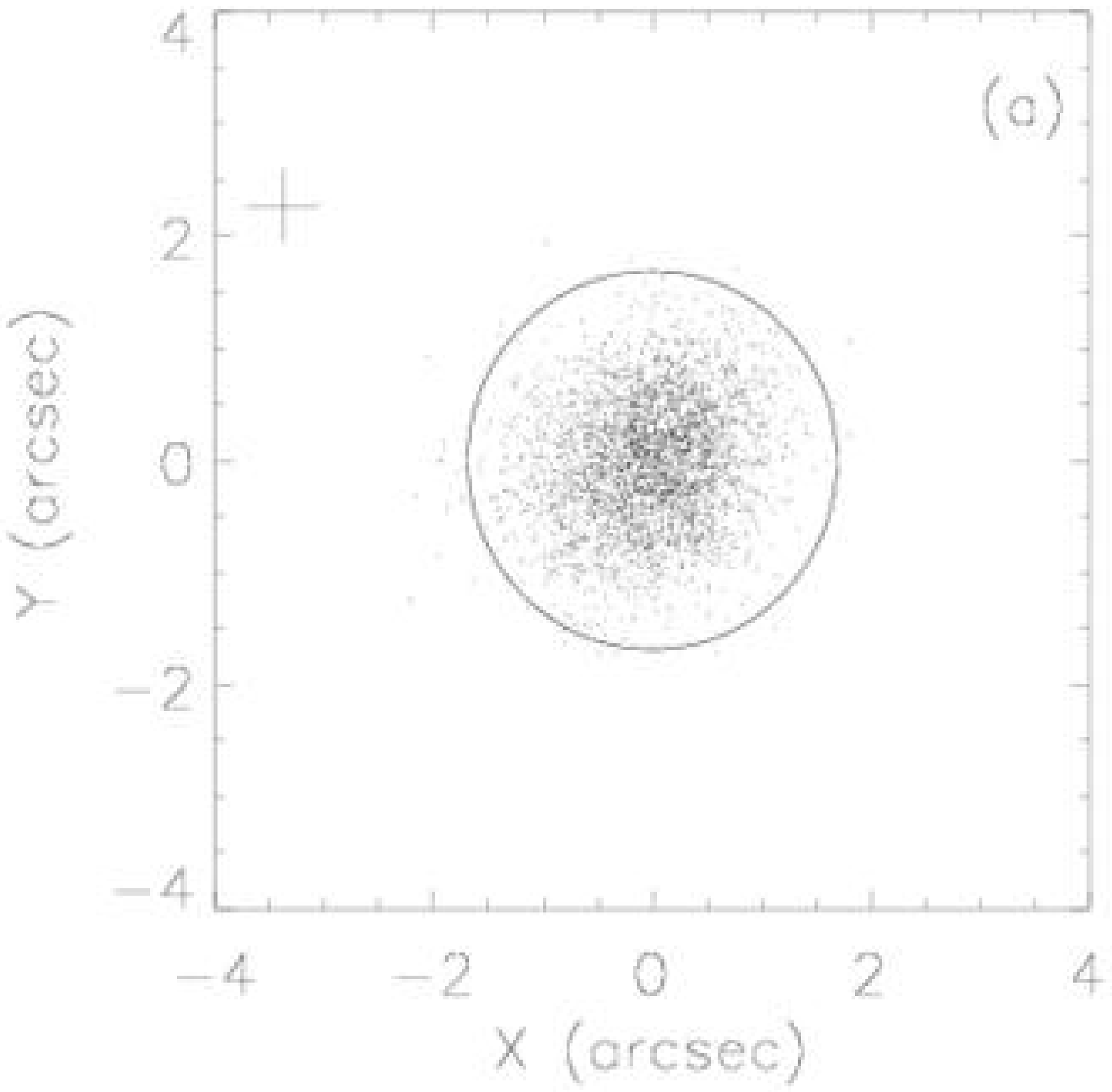}{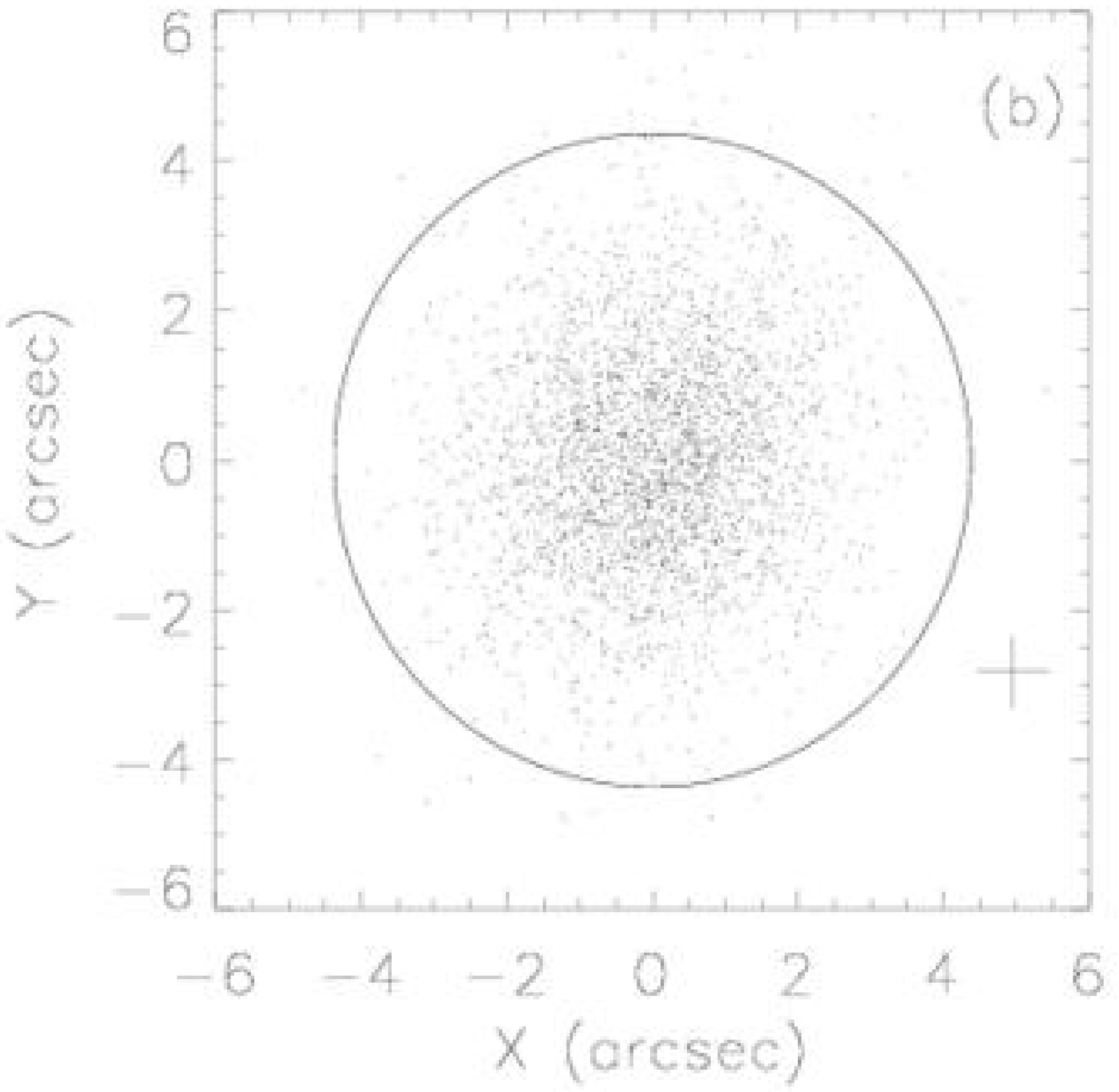}
\caption{Distribution of centroids in 5000 mass reconstructions. The luminosity peaks are marked by plus (+) symbol. In both
figures, the X-ray peaks are outside the plot range.
The circle represents the area where
99\% of the data is enclosed. No rejection is made. (a) Centroid distribution of the northern peak (clump C). The RMS is $\sim0.4 \arcsec$;
(b) Centroid distribution of the southern peak (clump F). The uncertainty increases to $\sim1.0\arcsec$.
\label{fig_centroid}}
\end{figure}

In our mass reconstruction of CL 0152-1357, we find that the two mass clumps detected in the weak lensing map are 
also shifted toward the suspected merging direction with respect to the luminosity as well as
the X-ray distribution (Figure~\ref{fig11.8} and~\ref{fig11.9}). 
Centroids of weak lensing mass 
clumps are in general affected by shape noise, shear strength, reconstruction algorithm artifacts, spatial number density  
variations of source galaxies, etc. Though most of these factors are present to a varying degree in our mass
reconstruction, we expect the uncertainties of these two centroids are relatively low compared
to those of outside the main body because of their high significance in the reconstructed mass map. Furthermore, it is
understood that the high number density of source galaxies ($\sim175$ objects $\mbox{arcmin}^{-2}$) increases
the overall stability of the centroids of our weak lensing mass peaks. One of the
useful tests to quantify the centroid uncertainty is to examine how much the locations of the mass peaks
change when the shears are perturbed slightly. Motivated by the experiment of \citet{cgm04}, we measured
the centroids of the two mass peaks in 5000 runs of mass reconstruction obtained in \textsection~\ref{bootstrap} by
bootstrap resampling. The resulting distribution of the two mass peaks are
presented in Figure~\ref{fig_centroid}. It shows that both of the two light peaks are outside the 99\% circle. The RMS
distance of the northen (clump C) and southern (clump F) peaks are $\sim0.4\arcsec$ and $\sim1.0\arcsec$, respectively. The smaller
RMS scatter of the northern peak is consistent with the higher significance in the reconstructed mass map.
The presence of these preferential shifts in the detected clumps may
extend the argument above and further support the collisionless nature of CDM.
Within the paradigm of the hierarchical structure formation, the shifts may indicate that the CDM which
initially created the deep potential well for the formation of the subclusters is moving
even faster than galaxies which are less subject to ram pressure than the ICM, but not entirely collionless as the CDM.

We observe that
there lie four bright foreground $(z\sim0.63)$ galaxies to the south of the BCGs. 
If they are massive enough to perturb the distortion of background galaxies, 
the centroid of the northern clump can be affected and appear to be further shifted toward the merging direction.
However, though we cannot completely rule out this possibility, we suspect that
their contribution to the centroid shift is not so substantive as to
cause the distinct offset between the X-ray and dark matter contours (Figure~\ref{fig11.9}).
More study of these questions regarding the centroid offsets among galaxies, X-ray emission, and weak lensing mass
will be conducted when 
the weak lensing analysis of another supposedly merging cluster, MS 1054-03 at $z\simeq0.83$,  
with ACS observations (M. Jee et al. 2004, in preparation) becomes available.

We compare the mass estimates of the cluster with those from the $Chandra$ X-ray spectral analysis \citep{maughan03}
and the Sunyaev-Zeldovich Effect (SZE) work \citep{joy01} by treating the cluster as a whole to simplify
the comparison.
\citet{joy01} infer that the total mass within a radius of $65\arcsec$ in $\Omega_M=0.3, \Omega_{\Lambda}=0.7,
$ and $H_0 = 100 $ $\mbox{km} \mbox{s}^{-1} \mbox{Mpc}^{-1}$ universe is $(2.1\pm0.7)\times10^{14} M_{\sun}$ from their SZE measurement of the cluster temperature.
The weak lensing mass under the same cosmological parameters yields $(1.9\pm0.2)\times10^{14} M_{\sun}$.
From the $Chandra$ X-ray analysis \citet{maughan03} quotes an estimate of $2.4_{-0.3}^{+0.4} \times 10^{14} M_{\sun}$ 
within $\sim50\arcsec$ radius aperture in $\Omega_M=0.3$, $\Omega_{\Lambda}=0.7$, and $H_0 = 70$ $\mbox{km} \mbox{s}^{-1} \mbox{Mpc}^{-1}$  universe.
Our conversion of the weak lensing mass under the same geometry is estimated to be $(2.1\pm0.3)\times10^{14} M_{\sun}$.
We note that the weak lensing mass estimate is lower by $\sim9$\% in both comparisons though 
the statistical significance of the difference is low. In the temperature-based measurement of the cluster mass, 
the dynamical equilibrium of the intracluster gas with the underlying dark matter as well as the spherical symmetry of
the matter distribution must be assumed. However, the on-going merger obviously violates these hypotheses and especially the
temperature rise caused by the shock may result in the overestimation of the cluster mass.

The X-ray mass estimate of $(1.1\pm0.2)\times10^{15} M_{\sun}$ within 1.4Mpc radius (extrapolated to the virial radius 
under the assumption of the isothermal sphere)
by \citet{maughan03} 
is roughly a factor of two higher than the result from the current paper even considering
the aforesaid risk of overestimation as well as the dissimilar geometry. We attribute this rather large difference
to the two following reasons. First, as discussed in \textsection\ref{summaryofmass}, the mass profile of CL 0152-1357
is better described by the NFW profile. 
The SIS modeling of the cluster profile gives
substantially higher total mass than the NFW representation does at large radii (leading to $\sim37$\% increase
at $\sim1.4$ Mpc). 
Second, the mass of the southern X-ray peak from this weak lensing analysis is much lower than the northern one
(less than 50 percent of the northern X-ray peak within $20\arcsec$ aperture radius, see Table~3) whereas
\citet{maughan03} estimates that the two peaks are of similar mass. Therefore, the total
virial mass of CL 0152-1357 computed by the superposition of two comparable SIS clumps is likely
to be much higher than the result from the current analysis.

\citet{huo04} presented the first weak lensing analysis of CL 0152-1357 using Keck R band observations. The
projected cluster mass of $\sim10^{15} M_{\sun}$ within $r=1 h_{65}^{-1}$ Mpc can be read off Figure 10 of 
their paper.
The transformation of our mass estimate in their cosmological parameters gives
$(5.09\pm0.46)\times10^{14} M_{\sun}$. Despite the somewhat large discrepancy in mass, we do
not further analyze the difference because the intermediate procedures of the weak lensing analysis (e.g.,
PSF corrections, tangential shears, reconstruction maps, redshift distribution of background galaxies, etc.) 
are not illustrated in their work.

\section{SUMMARY AND CONCLUSIONS\label{conclusion}}

We have presented our weak lensing analysis of the X-ray luminous cluster CL 0152-1357 at $z\sim0.84$
using ACS observations. The superb resolution and sensitivity of ACS provides high quality
images of weakly distorted, faint background galaxies in unprecedented depth. The resulting high number density of
source galaxies enables us to restore the cluster mass distribution in unparalleled detail when
the instrument artifact is properly accounted for. The complicated shape and variation
of the PSF is precisely modeled by exhaustive investigation of the 47 Tuc stars, and
the derived PSF is matched to the isolated good signal-to-ratio stars in CL 0152-1357 field after
a slight fine-tuning of the ellipticity. Rounding kernel test shows that the final
PSF obtained in this way nicely describes the observed PSF pattern in CL 0152-1357 field.
We use the publicly available GOODS and UDF images to infer the fraction of cluster galaxy contamination
in source galaxies, and the redshift distribution of background galaxies is estimated
by the use of the photometric redshift catalog of the UDF.

We determine the cluster mass via three different approaches: parameterized profile fitting, mass
reconstruction, and aperture mass densitometry. Among these approaches, the second method of
mass estimation from the reconstruction map is unconventional. The direct use of
the weak lensing mass map has been discouraged primarily because there exists
a sheet-mass degeneracy. However, we show that, after proper rescaling is considered, the method
yields very consistent results with the measurements from the often favored aperture mass densitometry.

Our weak lensing mass estimates at small radii ($r\lesssim65\arcsec$)
are consistent with the results from the X-ray emission and the Sunyaev-Zeldovich effect.
We show that the deviation of the mass profile from the
SIS profile increases at larger radii. 
The overall mass profile of the cluster can be well described by the NFW profile with
a scale radius of $r_s =309\pm45$ kpc and a concentration parameter of $c=3.7\pm0.5$. 
We estimate the total projected cluster mass within 
1 Mpc aperture to be (4.92 $\pm$ 0.44)$\times10^{14}M_{\sun}$ from the aperture mass densitometry.
The total luminosity of
the cluster is calculated by combining the spectroscopic and the red sequence catalogs.
When $i_{775}$ is transformed to the rest frame B, the total luminosity ($r \le 1$ Mpc) after accounting for
the blue and the faint population is determined
to be $L_{B\sun}=5.2 \times 10^{12} L_{B\sun}$. We find that the M/L ratio within 1 Mpc is
$95\pm$ 8 $M_{\sun} / L_{B\sun}$. Considering the luminosity evolution at $z\sim0.84$, this
M/L ratio corresponds to $\sim232 M_{\sun} / L_{B\sun}$ at local universe.

Our ACS weak lensing reveals very interesting substructure of the cluster in detail. The vertically
elongated cluster main body is clearly seen in both light and mass distributions with strong 
spatial correlations between light and mass clumps. Besides, we identify 4 scattered mass clumps outside the main body
with locations of cluster galaxy concentrations. More stimulating interpretation is made when
the mass reconstruction is compared with the X-ray morphology from $Chandra$ observations. In order
to examine the spatial correlations between the two analyses, we reprocess the $Chandra$ archival
data and overlay the X-ray contours with those of optical light and mass maps. We observe that the two
diffuse X-ray clumps are in spatial agreement with cluster galaxy concentrations, but are displaced
away from the assumed merging direction. The displacement of the southern peak was originally noticed by 
\citet{maughan03} and they suggested that the ICM is lagging behind the cluster galaxies due to
the ram pressure. The comparision of the X-ray emission with our mass reconstruction strengthens
this merger hypothesis because both the cluster galaxies and mass clumps seem to lead the X-ray
peaks. Furthermore, we remark on the displacements of the mass clumps relative to
the light concentrations. It appears in the mass-light comparison map
the two major mass clumps are slightly shifted ($\sim3 \arcsec$) toward the merging direction. 
The existence of these preferential
shifts might suggest that the collisionless dark matters are moving ahead of cluster galaxies.

Though the weak lensing survey data from today's
extensive dedication of many large aperture ground-based telescopes, primarily targeted for the cosmic
shear detection, surpass those of ACS in data volume, our weak lensing analysis of CL 0152-1357 
demonstrates that ACS is exclusively advantageous for weak lensing studies of high-redshift clusters 
which require only moderately large field of view, but extremely high resolution of the instrument.
The ACS GTO cluster survey program encompasses
many high-reshift clusters of great interest and the weak lensing investigation of these clusters
will provide many illuminating clues to the formation and evolution of galaxy clusters in the near future.

ACS was developed under NASA contract NAS 5-32865, and this research was supported
by NASA grant NAG5-7697. We are grateful for an equipment grant from the Sun Microsystems, Inc.

\clearpage

\clearpage

\clearpage

\clearpage

\clearpage

\clearpage 

\clearpage 

\clearpage

\clearpage

\clearpage

\clearpage

\clearpage 

\clearpage

\clearpage

\clearpage 

\clearpage

\clearpage

\clearpage 

\clearpage

\clearpage 

\clearpage 

\clearpage

\clearpage

\clearpage

\begin{deluxetable}{cccccccc} 
\tabletypesize{\scriptsize} 

\tablecaption{MASS PROPERTIES OF CL 0152-1357 (r $\le$ 1 Mpc)}

\tablenum{1}

\tablehead{\colhead{$\Omega_M$} & \colhead{$\Omega_{\Lambda}$} & \colhead{$H_0$} & 
\colhead{$M_{SIS}$} & \colhead{$M_{NFW}$} & \colhead{$M_{\zeta(r)}$} & \colhead{$M_{map}$} & \colhead{$M_{\zeta(r)}/L$} \\
\colhead{} & \colhead{} & \colhead{(km/s/Mpc)} 
 & \colhead{($10^{14}M_{\sun}$)} & \colhead{($10^{14}M_{\sun}$)} & \colhead{($10^{14}M_{\sun}$)} & \colhead{($10^{14}M_{\sun}$)} & \colhead{($M_{\sun}/L_{B\sun})$} \\  } 
\tablewidth{0pt}

\startdata
0.27 & 0.73 & 71 &  6.07 $\pm$ 0.75 &  4.92 $\pm$ 0.71 &  4.92 $\pm$ 0.44 &  4.74 $\pm$ 0.12 & 95 $\pm$ 8\\
1.0 & 0 & 50     &  7.31 $\pm$ 0.90 &  6.05 $\pm$ 0.88 &  5.94 $\pm$ 0.54 &  5.86 $\pm$ 0.12 & 108$\pm$10\\
\enddata
\end{deluxetable}

\begin{deluxetable}{cccccccc}
\tabletypesize{\scriptsize} 
\tablecaption{PROPERTIES OF SUBCLUMPS IN CL 0152-1357}

\tablenum{2}
\tablewidth{0pt}

\tablehead{\colhead{Subclump} & \colhead{$\zeta(\le20\arcsec)$} &  \colhead{$\bar{\kappa} (r_2 \le r \le r_{max})$} & \colhead{$r_2$} & \colhead{$r_{max}$} & 
\colhead{$M_{\zeta(r)}$} &
\colhead{$M_{map}$} & \colhead{$M_{map}/L$} \\ 
\colhead{} & \colhead{} & \colhead{} & \colhead{($\arcsec$)} & \colhead{($\arcsec$)} & \colhead{($10^{13} M_{\sun}$)} & \colhead{($10^{13} M_{\sun}$)} & 
\colhead{($M_{\sun}/L_{B\sun}$)} \\} 

\startdata
A & 0.064 $\pm$ 0.026 & 0.019 $\pm$ 0.013 & 40&60 & 2.2 $\pm$ 0.8 & 2.1 $\pm$ 0.3 & 73 $\pm$ 12\\
B & 0.120 $\pm$ 0.027 & 0.022 $\pm$ 0.006 & 80&100 & 3.7 $\pm$ 0.7 & 3.4 $\pm$ 0.2 & 80 $\pm$ 6\\
C & 0.232 $\pm$ 0.032 & 0.023 $\pm$ 0.003 & 140&160 & 6.2 $\pm$ 0.9 & 6.4 $\pm$ 0.2 & 123 $\pm$ 4\\
D & 0.117 $\pm$ 0.028 & 0.032 $\pm$ 0.006 & 80&100 & 3.9 $\pm$ 0.8 & 4.0 $\pm$ 0.2 & 215 $\pm$ 12\\
E & 0.100 $\pm$ 0.028 & 0.042 $\pm$ 0.006 & 80&100 & 3.8 $\pm$ 0.8 & 4.4 $\pm$ 0.2 & 148 $\pm$ 8\\
F & 0.068 $\pm$ 0.025 & 0.023 $\pm$ 0.016 & 60&80 & 2.4 $\pm$ 0.8 & 2.7 $\pm$ 0.2 & 61 $\pm$ 5\\
G & 0.064 $\pm$ 0.018 & 0.007 $\pm$ 0.018 & 40&60 & 1.9 $\pm$ 0.7 & 1.4 $\pm$ 0.3 & 106 $\pm$ 24\\
H & 0.076 $\pm$ 0.019 & 0.016 $\pm$ 0.020 & 40&60 & 2.4 $\pm$ 0.8 & 2.6 $\pm$ 0.3 & 120 $\pm$ 15\\
I & 0.052 $\pm$ 0.009 & 0.022 $\pm$ 0.020 & 25&35 & 1.9 $\pm$ 0.5 & 1.1 $\pm$ 0.3 & 174 $\pm$ 52\\
\enddata

\end{deluxetable}


\begin{thebibliography}{}
\bibitem[Bahcall \& Fan(1998)]{bf} Bahcall, N.~A.~\& Fan, X.\ 1998, \apj, 504, 1 
\bibitem[Bartelmann(1996)]{bartelmann96} Bartelmann, M.\ 1996, \aap,313, 697 
\bibitem[Bartelmann, Narayan, Seitz, \& Schneider(1996)]{bnss96} Bartelmann, M., Narayan, R., Seitz, S., \& Schneider, P.\ 1996, \apjl, 464, L115 
\bibitem[Beckwith, Somerville, \& Stiavelli(2003)]{beckwith03} Beckwith, S., Somerville, R., Stiavelli M., 2003, STScI Newsletter vol 20 issue 04 
\bibitem[Ben{\'{\i}}tez(2000)]{benitez00} Ben{\'{\i}}tez, N.\ 2000, \apj, 536, 571 
\bibitem[Ben{\'{\i}}tez et al.(2004)]{benitez04} Ben{\'{\i}}tez, N., et al.\ 2004, \apjs, 150, 
\bibitem[Bernstein \& Jarvis(2002)]{bj02} Bernstein, G.~M.~\& Jarvis, M.\ 2002, \aj, 123, 583 (BJ02)
\bibitem[Bertin \& Arnouts(1996)]{ba96} Bertin, E.~\& Arnouts, S.\ 1996, \aaps, 117, 393 
\bibitem[Blakeslee et al.(2003)]{blakeslee03} Blakeslee, J.~P., Anderson, K.~R., Meurer, G.~R., Ben{\'{\i}}tez, N., \& Magee, D.\ 2003, ASP 
Conf.~Ser.~295: Astronomical Data Analysis Software and Systems XII, 12, 257 
\bibitem[Bonnet et al.(1993)]{bonnet93} Bonnet,H.,Fort,B.,Kneib,J.-P.,Mellier,Y.,\& Soucail,G. 1993,\aap, 280, L7
\bibitem[Bonnet, Mellier, \& Fort,(1994)]{bonnet94} Bonnet,H.\&Mellier,Y.,\& Fort,B. 1994,\apjl, 427, L83   
\bibitem[Bower \& Smail (1997)]{bs97} Bower,R.G. \& Smail, I. 1997, \mnras, 290, 292   
\bibitem[Broadhurst, Taylor, \& Peacock (1995)]{btp95} Broadhurst, T., Taylor, A.N., \& Peacock, J.A. 1995,\apj, 438, 49   
\bibitem[Carlberg, Morris, Yee, \& Ellingson(1997)]{cmye} Carlberg, R.~G., Morris, S.~L., Yee, H.~K.~C., \& Ellingson, E.\ 1997, \apjl, 479, L19 
\bibitem[Carlberg, Yee, \& Ellingson(1997)]{cye97} Carlberg, R.~G., Yee, H.~K.~C., \& Ellingson, E.\ 1997, \apj, 478, 462 
\bibitem[Clowe et al.(1998)]{clowe98} Clowe, D., Luppino, G.~A., Kaiser, N., Henry, J.~P., \& Gioia, I.~M.\ 1998, \apjl, 497, L61 
\bibitem[Clowe, Gonzalez, \& Markevitch(2004)]{cgm04} Clowe, D., Gonzalez, A., \& Markevitch, M.\ 2004, \apj, 604, 596 
\bibitem[Coleman, Wu, \& Weedman(1980)]{cww80} Coleman, G.~D., Wu, C.-C., \& Weedman, D.~W.\ 1980, \apjs, 43, 393 
\bibitem[Della Ceca, Maccacaro, Rosati, \& Braito(2000)]{della00} Della Ceca, R., Maccacaro, T., Rosati, P., \& Braito, V.\ 2000, \aap, 355, 121 
\bibitem[Ebeling et al.(2000)]{ebeling00} Ebeling, H., et al.\ 2000, \apj, 534, 133 
\bibitem[Fahlman et al.(1994)]{fahlman94} Fahlman,G.,Kaiser,N.,Squires,G.\& Woods,D. 1994,\apj, 437, 56
\bibitem[Fischer et al.(1997)]{fischer} Fischer, P., Bernstein, G., Rhee, G., \& Tyson, J.A. 1997,\aj, 113, 521
\bibitem[Fischer \& Tyson(1997)]{ft97} Fischer, P.~\& Tyson, J.~A.\ 1997, \aj, 114, 14 
\bibitem[Fort \& Mellier (1994)]{fm94} Fort, B., Mellier, Y., Dantel-Fort, M., Bonnet, H., \& Kneib, J.-P. 1996,\aap, 310, 705
\bibitem[Fort et al. (1996)]{fort96} Fort,B.\& Mellier,Y. 1994,\aapr,5,239
\bibitem[Giavalisco et al.(2004)]{giavalisco04} Giavalisco, M., et al.\ 2004, \apjl, 600, L93
\bibitem[Hirata \& Seljak(2003)]{hs03} Hirata, C.~\& Seljak, U.\ 2003, \mnras, 343, 459 
\bibitem[Hoekstra, Franx, \& Kuijken(2000)]{hfk00} Hoekstra, H., Franx, M., \& Kuijken, K. 2000,\apj, 532, 88 (HFK00)
\bibitem[Huo et al.(2004)]{huo04} Huo, Z., Xue, S., Xu, H., Squires, G., \& Rosati, P.\ 2004,\aj, 127, 1263
\bibitem[Joy et al.(2001)]{joy01} Joy, M., et al.\ 2001, \apjl, 551, L1 
\bibitem[Kaiser \& Squires(1993)]{ks93} Kaiser, N.~\& Squires, G.\ 1993, \apj, 404, 441 
\bibitem[Kaiser, Squires, \& Broadhurst(1995)]{ksb95} Kaiser, N., Squires, G., \& Broadhurst, T. 1995,\apj, 449, 460
\bibitem[Kaiser(2000)]{kaiser00} Kaiser, N.\ 2000, \apj, 537, 555 
\bibitem[Katgert, Biviano, \& Mazure(2004)]{kbm04} Katgert, P., Biviano, A., \& Mazure, A.\ 2004, \apj, 600, 657 
\bibitem[Kinney et al.(1996)]{kinney96} Kinney, A.~L., Calzetti, D., Bohlin, R.~C., McQuade, K., Storchi-Bergmann, T., \& Schmitt, H.~R.\ 1996, \apj, 467, 38 
\bibitem[King \& Schneider(2001)]{ks01} King, L.~J.~\& Schneider, P.\ 2001, \aap, 369, 1 
\bibitem[Kochanek(1990)]{kochanek90} Kochanek, C.S. 1990, \mnras, 247, 135
\bibitem[Kormann, Schneider, \& Bartelmann(1994)]{ksb94} Kormann, R., Schneider, P., \& Bartelmann, M.\ 1994, \aap, 284, 285 
\bibitem[Krist(2003)]{krist03} Krist, J.\ 2003, $Instrument$ $Science$ $Report$ $ACS$ 2003-06
\bibitem[Kuijken(1999)]{kuijken99} Kuijken, K.\ 1999, \aap, 352, 355 
\bibitem[Lecar(1975)]{lecar75} Lecar, M.\ 1975, IAU Symp.~ 69: Dynamics of the Solar Systems, 69, 161 
\bibitem[Lombardi \& Bertin(1999)]{lb99} Lombardi, M.~\& Bertin, G.\ 1999, \aap, 348, 38 
\bibitem[Markevitch et al.(2002)]{markevitch02} Markevitch, M.,Gonzalez, A.~H., David, L., Vikhlinin, A., Murray, S., Forman, W., Jones, C., \& Tucker, W.\ 2002, \apjl, 567, L27 
\bibitem[Marshall, Hobson, Gull, \& Bridle(2002)]{mhgb02} Marshall, P.~J., Hobson, M.~P., Gull, S.~F., \& Bridle, S.~L.\ 2002, \mnras, 335, 1037 
\bibitem[Maughan et al.(2003)]{maughan03} Maughan, B.~J., Jones, L.~R., Ebeling, H., Perlman, E., Rosati, P., Frye, C., \& Mullis, C.~R.\ 2003, \apj, 587, 589 
\bibitem[McCann(2004)]{mccann04} McCann, W.J. 2004, http://acs.pha.jhu.edu/general/software/fitscut/
\bibitem[Mellier et al.(1994)]{mellier94} Mellier,Y.,Dantel-Fort,M.,Fort,B.,\& Bonnet,H. 1994,\aap, 289, L15
\bibitem[Merritt(1988)]{merritt88} Merritt, D.\ 1988, ASP Conf.~Ser.~  5: The Minnesota lectures on Clusters of Galaxies and Large-Scale Structure, 175 
\bibitem[Meurer et al.(2003)]{meurer03} Meurer, G.~R., et al.\ 2003, \procspie, 4854, 507 
\bibitem[Miralda-Escud\'e(1991)]{miralda91} Miralda-Escud\'e,J. 1991,\apj,380,1
\bibitem[Mobasher et al.(2004)]{mobasher04} Mobasher, B., et al.\ 2004, \apjl, 600, L167 
\bibitem[Navarro, Frenk, \& White(1997)]{nfw97} Navarro, J.~F., Frenk, C.~S., \& White, S.~D.~M.\ 1997, \apj, 490, 493 
\bibitem[Refregier(2003)]{refregier03} Refregier, A.\ 2003, \mnras, 338, 35 (R03)
\bibitem[Richmond(2002)]{richmond02} Richmond, M. 2002, http://acd188a-005.rit.edu/match/
\bibitem[Romer et al.(2000)]{romer00} Romer, A.~K., et al.\ 2000, \apjs, 126, 209 
\bibitem[Rosati, della Ceca, Norman, \& Giacconi(1998)]{rosati98} Rosati, P., della Ceca, R., Norman, C., \& Giacconi, R.\ 1998, \apjl, 492, L21 
\bibitem[Sarazin(1988)]{sarazin88} Sarazin, C.~L.\ 1988, \skytel, 76, 639 
\bibitem[Scharf et al.(1997)]{scharf97} Scharf, C.~A., Jones, L.~R., Ebeling, H., Perlman, E., Malkan, M., \& Wegner, G.\ 1997, \apj, 477, 79 
\bibitem[Schlegel, Finkbeiner, \& Davis(1998)]{sfd98} Schlegel, D.~J., Finkbeiner, D.~P., \& Davis, M.\ 1998, \apj, 500, 525 
\bibitem[Schneider \& Seitz(1995)]{ss95a} Schneider, P.~\& Seitz, C.\ 1995, \aap, 294, 411 
\bibitem[Schneider, King, \& Erben(2000)]{ske00} Schneider, P., King, L., \& Erben, T.\ 2000, \aap, 353, 41 
\bibitem[Seitz \& Schneider(1995)]{ss95b} Seitz, C.~\& Schneider, P.\ 1995, \aap, 297, 287 
\bibitem[Seitz et al.(1996)]{seitz96} Seitz, C., Kneib, J.-P., Schneider, P., Seitz, S. 1996,\aap, 314, 707
\bibitem[Seitz \& Schneider(1997)]{ss97} Seitz, C.~\& Schneider, P.\ 1997, \aap, 318, 687 
\bibitem[Seitz, Schneider, \& Bartelmann(1998)]{seitz98} Seitz, S., Schneider, P., \& Bartelmann, M.\ 1998, \aap, 337, 325 
\bibitem[Seitz \& Schneider(2001)]{ss01} Seitz, S.~\& Schneider, P.\ 2001, \aap, 374, 740 
\bibitem[Smail \& Dickinson (1995)]{sd95} Smail,I. \& Dickinson,M. 1995,\apjl, 455, L99
\bibitem[Smail et al.(1997)]{smail97} Smail, I., Ivison, R.J., \& Blain,A.W. 1997,\apjl, 490, L5
\bibitem[Squires \& Kaiser(1996)]{sk96} Squires, G.~\& Kaiser, N.\ 1996, \apj, 473, 65 
\bibitem[Takahashi, sensui, Funato, \& Makino(2002)]{tsfm02} Takahashi, K., sensui, T., Funato, Y., \& Makino, J.\ 2002, \pasj, 54, 5 
\bibitem[Tyson \& Fisher (1995)]{tf95} Tyson,J.A. \& Fischer, P. 1995,\apjl, 446, L55
\bibitem[Tyson, Wenk, \& Valdes(1990)]{twv} Tyson, J.~A.,Wenk, R.~A., \& Valdes, F.\ 1990, \apjl, 349, L1 
\bibitem[van Dokkum \& Stanford (2003)]{vs03} van Dokkum, P.~G.~\& Stanford, S.~A.\ 2003, \apj, 585, 78
\bibitem[Wilson, Cole, \& Frenk(1996)]{wcf96} Wilson, G., Cole, S., \& Frenk, C.~S.\ 1996, \mnras, 282, 501 
\bibitem[Wright \& Brainerd(2000)]{wb00} Wright, C.~O.~\& Brainerd, T.~G.\ 2000, \apj, 534, 34 
\bibitem[Wu et al.(1998)]{wu98} Wu, X., Chiueh, T., Fang, L., \& Xue, Y.\ 1998, \mnras, 301, 861


\end{thebibliography}
\end{document}